\documentclass[letterpaper]{JHEP3}

\usepackage{epsf,epsfig}
\usepackage{amssymb}
\usepackage{graphicx}
\usepackage{amsmath}
\usepackage{amsfonts}

\usepackage{graphicx}

\newcommand{\insertplot}[5]{\begin{figure}
 \hfill\hbox to 0.05in{\vbox to #5in{\vfill
 \inputplot{#1}{#4}{#5}}\hfill}
 \hfill\vspace{-.1in}
 \caption{#2}\label{#3}
 \end{figure}}
 \newcommand{\inputplot}[3]{
 \special{ps: plotfile #1}
}
\newcounter{fig}

\newcommand{\beq}{\begin{equation}}
\newcommand{\eeq}{\end{equation}}
\newcommand{\beqs}{\begin{eqnarray}}
\newcommand{\eeqs}{\end{eqnarray}}

\newcommand{\be}{\begin{equation}}
\newcommand{\ee}{\end{equation}}
\newcommand{\bea}{\begin{eqnarray}}
\newcommand{\eea}{\end{eqnarray}}

\numberwithin{equation}{section}

\abstract{ 
 The five dimensional Einstein-Gauss-Bonnet gravity with a
 negative cosmological constant  
 becomes,  for a special value of the Gauss-Bonnet coupling constant,
a Chern-Simons (CS) theory of gravity.
In this work we discuss the properties of 
several different types of black object solutions
of this model.
Special attention is paid to the case of spinning black holes with equal-magnitude angular momenta 
which
 posses  a regular horizon of spherical topology. 
Closed form solutions are obtained in the small angular momentum limit. 
 Nonperturbative solutions are constructed by solving numerically the equations of the model.
Apart from that,  new exact solutions describing static squashed black holes 
  and  black strings
are also discussed.
The action and global charges of all configurations studied in this work are
obtained by using the quasilocal formalism 
with boundary counterterms generalized 
for the case of a $d=5$ CS theory. 

 }

\keywords{Chern-Simons theory of gravity, black holes, numerical solutions}
\preprint{ }

\title{Black hole solutions in $d=5$ Chern-Simons gravity }

\author{{\large Yves Brihaye}$^{\dagger}$  
and {\large Eugen Radu}$^{\ddagger }$ \\ \\
$^{\dagger}${\small Physique-Math\'ematique, Universite de
Mons-Hainaut, Mons, Belgium}\\
$^{\ddagger}${\small Institut f\"ur Physik, Universit\"at Oldenburg, Postfach 2503
D-26111 Oldenburg, Germany}

  }

\begin{document}

\section{Introduction}
It is known that in dimensions higher than four, the Einstein theory of gravity can 
be generalized by including in the action a linear combination of all   
lower dimensional Euler densities  \cite{Lovelock:1971yv}.
This gives rise to the so-called Lovelock models of gravity
which
possess some special features.
For example,  the field
equations still obey generalized Bianchi identities and are linear in the 
second order derivatives of the metric (this being also the maximal order).
Moreover, the  Lovelock models are known to be free of ghosts
when expanding around a flat space background, 
thus avoiding problems with unitarity 
\cite{Zwiebach:1985uq}, 
\cite{Boulware:1985wk}.

However, as first discussed in \cite{Chamseddine:1989nu},
in odd spacetime dimensions there exist a special combination of the terms which enter the
corresponding
Lovelock gravity model 
that allows to write the corresponding Lagrangeans as Chern-Simons (CS) densities.
The resulting models retain all properties of the Lovelock 
gravity, exhibiting at the same time some new interesting features.
For example, the CS theory possesses
an enhanced local symmetry and can be reformulated as a gauge theory of gravity.

In this work we shall restrict ourselves to the case of a CS theory 
in $d=5$ spacetime dimensions with a negative cosmological constant.
In this case, the Lovelock model of gravity corresponds to the so-called 
Einstein-Gauss-Bonnet (EGB) theory, which contains quadratic powers of 
the curvature. 
As discussed by many authors, 
the inclusion of a GB term in the gravity action leads to a variety of 
new features (see  \cite{Garraffo:2008hu}, \cite{Charmousis:2008kc} for reviews 
of these models in the larger context of higher order gravity theories).
In particular, the  black holes of EGB theory do not in general obey the
Bekenstein-Hawking area law,
but the entropy formula includes a new
contribution coming from the higher curvature terms in the action 
\cite{Jacobson:1993xs,Wald:1993nt}.

The $d=5$ solutions of the EGB equations with
an Anti-de Sitter (AdS)  spacetime background
have been extensively studied in the recent years,
 mainly motivated
 by the conjectured AdS/CFT 
 correspondence\footnote{In this framework, the introduction of higher order terms 
 in the gravity action corresponds to next to leading order corrections to 
the $1/N$ expansion of the boundary dual theory 
\cite{Fayyazuddin:1998fb}, 
\cite{Aharony:1998xz}, 
\cite{Nojiri:2000gv}.} \cite{Maldacena:1997re}.  
Then a set of the EGB-AdS solutions with special a value of
the GB coupling constant $\alpha$ fixed by the cosmological constant $\Lambda$, 
corresponds also to solutions of the CS model. 
However, a search in the literature shows that 
the only solutions of the CS model
which have been extensively studied are the
  counterparts of the 
Schwarzschild black holes
(see $e.g.$ 
\cite{Cai:1998vy}, 
\cite{Crisostomo:2000bb}, 
\cite{Aiello:2004rz}).
Moreover, several classes of solutions 
which are known to exist in  Einstein gravity 
($e.g.$ black strings and spinning black holes)
are still missing in this case. 
Also, at a technical level, the solutions of the CS model are rather special, 
due to the fact that for the specific ratio between the $\alpha$
and $\Lambda$, the equations of motion of the Lovelock theory
become somehow degenerate. 
As we shall see, this leads to the existence of exact solutions
in a number of cases where no closed form solutions could be found 
in the EGB case.
Moreover, 
some results found in the generic EGB model cannot be safely 
extrapolated to the particular case of a CS model. 

The main purpose of this work is to provide a discussion of
several different classes of solutions of the $d=5$ CS model, looking for generic properties.
In the static case, we consider 
generalizations of the squashed  Schwarzschild-AdS black holes in 
\cite{Murata:2009jt}, 
\cite{Brihaye:2009dm},
as well as the CS counterparts  of the
asymptotically AdS black strings in 
\cite{Copsey:2006br}, \cite{Mann:2006yi}.
Different from the case of pure Einstein gravity,
we are able to find in these case a simple analytic
expression of the solutions. 

Apart from these static solutions,  
we consider also rotating black holes  
with equal magnitude angular momenta and possessing a spherical horizon topology.
Since in this case we could not 
find closed form solutions (except in the slowly rotating limit),
these configurations are constructed numerically 
by matching the near horizon expansion of the metrics
to their asymptotic form.

The paper is organized as follows.
In the next Section we explain 
the CS model and describe 
the computation of the physical quantities of the solutions such as 
their action and  mass-energy. 
Our proposal for the general counterterm expression in $d=5$ CS theory 
is also presented there. 
The next two Sections contains applications of the general formalism.
In  Section 3 we give our results for the static black holes and black strings solutions.
Then in Section 4 we discuss the basic properties of the spinning black hole solutions
with an $S^3$ event horizon topology.
We conclude in Section 5
with some further remarks.
The Appendices contain technical details on the rotating black holes
together with an exact solution describing a singular spinning configuration.

\section{The model}

\subsection{The action}
We start by considering 
the five dimensional Einstein-Hilbert action with a negative cosmological constant,
 supplemented by quadratic terms:
\begin{eqnarray}
\label{EGB0}
I = \frac{1}{16\pi G}\int_{ \mathcal{M}} \sqrt{-g}\bigg[ R-2\Lambda
+ \frac{1}{4}
\left(
\alpha_1 R^2 + \alpha_2 R_{\mu \nu}R^{\mu \nu} +\alpha_3 R_{\mu \sigma \kappa \tau}R^{\mu \sigma \kappa \tau}
\right)
\bigg] 
d^5 x, 
\end{eqnarray}
where 
$g$ is the determinant of the metric tensor $g_{\mu\nu}$, $R$ is the Ricci scalar,  
$R_{\mu \nu}$ is the Ricci tensor, 
$R_{\mu \sigma \kappa \tau}$ is the Riemann tensor and
 $\Lambda=-6/\ell^2$ is the cosmological constant.

 The case 
  \be
  \label{cond}
\alpha_1=-\frac{1}{4}\alpha_2=\alpha_1=\alpha
 \ee
 is special, 
 the quadratic part in (\ref{EGB0})
  becoming the Lagrangeean of the Gauss-Bonnet gravity,
 \be
 L_{GB}=R^2 - 4 R_{\mu \nu}R^{\mu \nu} + R_{\mu \sigma \kappa \tau}R^{\mu \sigma \kappa \tau} ~.
 \ee
The constant $\alpha$ in (\ref{cond}) is the GB coefficient with dimension $(length)^2$ and is positive
in the string theory. 
For the case of a CS theory in this work, its value is fixed by the cosmological constant, 
\begin{eqnarray}
\label{alpha}
\alpha=-\frac{3}{\Lambda}=\frac{\ell^2}{2}.
\end{eqnarray}
Then, as first discussed in \cite{Chamseddine:1989nu}, the resulting model can 
 be thought as a higher-dimensional generalization of the well-known CS
formulation of the three-dimensional Einstein gravity \cite{Witten:1988hc}.
 Without entering into details, we mention that to show this construction
 explicitly, it is necessary to employ the first
 order formalism in terms of the spin connection $w^{ab}=w_{\mu}^{ab}dx^\mu$
 with the veilbeins $e^a_\mu$ which define the metric $g_{\mu\nu}=\eta_{ab}e^a_\mu e^b_\nu$
 and the one forms $e^a=e^a_\mu dx^\mu$.
 Then $R^{ab}=dw^{ab}+w^a_cw^{cb}$  is the curvature two form 
 (where the wedge product between forms is understood)
 and (\ref{EGB0}) can be written as a CS form for the AdS group \cite{Chamseddine:1989nu}
\begin{eqnarray}
 \label{CS5}
I = \frac{1}{16\pi G}\int_{ \mathcal{M}} 
Tr
\left \{
AdAdA+\frac{3}{2}dA A^3+\frac{3}{5}A^5
\right \}
,
\end{eqnarray}
 where
 \begin{eqnarray}
A^{ab}=%
\begin{pmatrix}
\omega ^{ab} & e^{a}/l \\ 
-e^{b}/l & 0%
\end{pmatrix}%
,
\end{eqnarray}
 with $a,b$ running here from 1 to 6
(thus $A^{ab}$ is a six-dimensional one form).
One should mention that different from the three dimensional CS gravity theory \cite{Witten:1988hc}
which has no propagating degrees of freedom, 
the model (\ref{CS5}) is a fully interacting theory.
Also, it is clear that this theory does not present a limit where
the standard general relativity
is recovered.
The corresponding supergravity generalizations have been constructed in 
\cite{Chamseddine:1990gk},
\cite{Banados:1996hi},
\cite{Troncoso:1997va} (see \cite{Zanelli:2005sa} for an extensive 
introduction to the  subject of CS gravity and supergravity).

\subsection{Field equations and boundary terms}

For the purposes of this work, however, it is more convenient to use the usual formulation
of the model in a coordinate basis. 
Then the variation of the  (\ref{EGB0})  
with respect to the metric
tensor $g_{\mu\nu}$ 
(with the choices (\ref{cond}), (\ref{alpha}) of the coupling constants) results in the field equations 
\begin{eqnarray}
\label{eqs}
&&
E_{\mu \nu } \equiv R_{\mu \nu } -\frac{1}{2}Rg_{\mu \nu}
 -\frac{6}{\ell^2}  g_{\mu \nu }
 \\
 \nonumber
&&
 {~~~~~~~~}
+\frac{\ell^2}{8}\bigg(
 2(R_{\mu \sigma \kappa \tau }R_{\nu }^{\phantom{\nu}%
\sigma \kappa \tau }
-2R_{\mu \rho \nu \sigma }R^{\rho \sigma }
-2R_{\mu
\sigma }R_{\phantom{\sigma}\nu }^{\sigma }+RR_{\mu \nu })-\frac{1}{2}%
L_{GB}g_{\mu \nu } 
 \bigg)
 =0.
\end{eqnarray}
In this work we are interested in solutions of (\ref{eqs}) approaching asymptotically a spacetime  of negative constant curvature.
This implies the asymptotic
expression of the Riemann tensor\footnote{A  precise definition 
of asymptotically AdS spacetime in higher curvature gravitational theories 
is given $e.g.$ in \cite{Okuyama:2005fg}.}:
$R_{\mu \nu}^{~~\lambda \sigma}=-(\delta_\mu^\lambda \delta_\nu^\sigma
-\delta_\mu^\sigma \delta_\nu^\lambda)/\ell_{eff}^2$,
where $\ell_{eff}=\ell/\sqrt{2}$ is the new effective radius of the AdS space in CS theory.
It is worth mentioning that this 
asymptotically locally 
AdS condition reflects
a local property at the boundary, but it does not
restrict the global topology of the spacetime manifold.
In fact, there is a wide class of solutions that satisfy this
condition, including the (squashed) black holes and black strings in this work.
Also, similar to the pure Einstein gravity case,
the  solutions can be written in Gauss-normal coordinates 
and admit a Fefferman-Graham--like asymptotic expansion \cite{Feff-Graham}.

Returning to the issue of model's action, we mention that
for a well-defined variational principle with a fixed metric on the boundary, 
one has to supplement (\ref{EGB0}) with 
a boundary term  which is the sum of the 
the Gibbons-Hawking surface term \cite {GibbonsHawking1} 
and its counterpart for GB gravity  \cite{Myers:1987yn}:
\begin{equation}
\label{Ib1}
I_{b}=-\frac{1}{8\pi G}\int_{\partial \mathcal{M}}d^{4}x\sqrt{-\gamma }
\left(
K
+\frac{\ell^2}{4} ( J-2\ {G}_{ab} K^{ab} )
\right)
.
\end{equation}
 In the above relation, $\gamma _{ab }$ is the induced metric on the boundary,    
$K_{ab}$  is the extrinsic curvature tensor of the boundary
 and $K$ the 
trace of this tensor,
  $\ {G}_{ab}$ is the Einstein tensor of the metric $\gamma _{ab}$ and $J$ is the
trace of the tensor
\begin{equation}
J_{ab}=\frac{1}{3}%
(2KK_{ac}K_{b}^{c}+K_{cd}K^{cd}K_{ab}-2K_{ac}K^{cd}K_{db}-K^{2}K_{ab})~.
\label{Jab}
\end{equation}

\subsection{The counterterms and boundary stress tensor}
The action and global charges of the solutions  in this work 
are computed by using
the procedure proposed by Balasubramanian and Kraus 
\cite{Balasubramanian:1999re}  for the case of the Einstein gravity
with negative cosmological constant. This technique was inspired by the 
AdS/CFT correspondence (since quantum field theories in general contains 
counterterms)
and consists of adding to the action suitable 
boundary counterterms $I_{ct}$, which are functionals only of 
curvature invariants of the induced metric on the boundary.  
These counterterms remove all power-law divergencies from the on-shell action,
without being necessary to specify a background metric.

In this approach, we supplement the general action 
(which consist in  (\ref{EGB0}) together with the boundary term (\ref{Ib1}))  
with  the following boundary counterterm\footnote{A different construction of the holographic
stress tensor in CS gravity
can be found in 
\cite{Banados:2004zt},
\cite{Banados:2005rz},
together with a discussion of the holographic anomalies.
}
\begin{eqnarray}
\label{Lagrangianct} 
I_{\mathrm{ct}}  &=&
-\frac{1}{8\pi G}
\int_{\partial \mathcal{M}} d^4x
\sqrt{-\gamma}
\bigg (
 \frac{2\sqrt{2}}{\ell}
  +\frac{\ell} {2\sqrt{2}} \mathsf{R}
 \bigg),
\end{eqnarray}
where $\mathsf{R}$ and $\mathsf{R}^{ab}$   are the curvature and the 
Ricci tensor  associated with the induced metric $\gamma $.  
The expression  (\ref{Lagrangianct}) is obtained by taking the limit $\alpha\to \ell^2/2$
in the general EGB counterterm expression proposed
 in \cite{Brihaye:2008kh}, \cite{Brihaye:2008xu}.
 
 Then one can define a boundary stress-energy tensor
 which is the variation of the total action with respect to the boundary metric, 
$
T_{ab}=\frac{2}{\sqrt{-\gamma }}\frac{\delta }{\delta \gamma ^{ab}}\left(
I+I_{b}+I_{\text{ct}} \right)  
$.
Its explicit expression is
\begin{eqnarray}
 T_{ab}&=&
 \frac{1}{8\pi G}
 \left [
K_{ab}-\gamma _{ab}K
 +\frac{{\ell^2}}{4} (Q_{ab}-\frac{1}{3}Q\gamma_{ab}) 
-\frac{2\sqrt{2}}{\ell}\gamma _{ab} 
+\frac{\ell}{\sqrt{2}}
\left( \mathsf{R}_{ab}-\frac{1}{2}\gamma _{ab}\mathsf{R} \right)
\right ]
,  
\end{eqnarray} 
where we define
\begin{eqnarray}
&&
Q_{ab}= 
2KK_{ac}K^c_b-2 K_{ac}K^{cd}K_{db}+K_{ab}(K_{cd}K^{cd}-K^2)
\\
\nonumber
&&
{~~~~~~}
+2K \mathsf{R}_{ab}+\mathsf{R}K_{ab}
-2K^{cd}\mathsf{ R}_{cadb}-4 \mathsf{R}_{ac}K^c_b.~~
\end{eqnarray} 
The computation of the global charges associated with the Killing symmetries of
the boundary metric is done in a similar way to the well-known 
case of Einstein gravity \cite{Balasubramanian:1999re}.
The boundary submanifold of all solutions in this work
can be foliated in a standard ADM  form, with
$
\gamma_{ab}dx^adx^b=-N^2dt^2+\sigma_{ij}(dy^i+N^idt)(dy^j+N^jdt),
$
where $N$ and $N^i$ are the lapse function, respectively 
the shift vector, and $y^i$ are the intrinsic 
coordinates on a closed surface $\Sigma$ of constant time 
$t$ on the boundary.  
Then a conserved charge 
\begin{equation}
{\mathfrak Q}_{\xi }=\oint_{\Sigma }d^{3}y\sqrt{\sigma}u^{a}\xi ^{b}T_{ab},
\label{Mcons}
\end{equation}%
can be associated with the closed surface $\Sigma $ 
(with normal $u^{a}$), provided the boundary geometry 
has an isometry generated by a Killing vector $\xi ^{a}$. 
For example,  the conserved mass/energy ${\cal M}$ is the charge associated 
with the time translation symmetry (with  $\xi =\partial /\partial t$),
being computed from the $tt$-component of the boundary stress tensor. 

Once the global charges are computed, the thermodynamics
of the vacuum black objects in this work is formulated by using the standard Euclidean approach
\cite{Hawking:ig}.
 In the semiclassical approximation, the partition function is given by $Z\simeq e^{-I_{cl}}$,
 where $I_{cl}$ is the classical action ($i.e.$ the sum of $I,~I_b$ and $I_{ct}$) evaluated on the equations of motion.   
As usual, the Hawking 
temperature $T_H$ of a black object  is found by demanding regularity of the 
Euclideanized manifold, or equivalently,  by evaluating the surface gravity.  
Upon application of the Gibbs-Duhem relation to the partition 
function  (see $e.g.$ \cite{Mann:2003-Found}),
this yields an expression for the entropy (with $\beta=1/T_H$)
\begin{equation}
S=\beta ({\cal M}-\mu _{i}{\mathfrak C}_{i})-I_{cl},  
\label{GibbsDuhem}
\end{equation}%
 with chemical potentials ${\mathfrak C}_{i}$ and
conserved charges\footnote{The expression of ${\mathfrak C}_{i},\mu _{i}$ depends on the physical situation. 
For example, the spinning black holes in Section 4 have ${\mathfrak C}=\Omega_{1,2}=\Omega_H$ and $\mu=J_{1,2}=J$. } 
$\mu _{i}$.
The entropy  
can also be written in Wald's form \cite{Wald:1993nt}
as an integral over the event horizon 
\begin{eqnarray}
\label{S-Noether} 
S=\frac{1}{4G}\int_{\Sigma_h} d^{3}x \sqrt{\tilde h}(1+\frac{\ell^2}{4}\tilde R),
\end{eqnarray} 
(where $\tilde h$ is the determinant of the induced metric on the horizon 
and $\tilde R$ is the event horizon curvature). 
As usual, the first law of thermodynamics  
\begin{equation}
dS=\beta (d{\cal M}-\mu _{i}d{\mathfrak C}_{i}). 
\label{1stlaw}
\end{equation}
provides an important test of the consistency of results.

The background metric upon which the dual field theory resides is defined as
$h_{ab}=\lim_{r \rightarrow \infty} \frac{\ell^2}{2r^2}\gamma_{ab}$.
The expectation value of the stress tensor of the dual theory 
can be computed using the  relation 
\cite{Myers:1999ps}:
\begin{eqnarray} 
\label{Tik-CFT}
\sqrt{-h}h^{ab}<\tau _{bc}>=\lim_{r\rightarrow \infty }\sqrt{-\gamma 
}\gamma
^{ab}T_{bc}.
\end{eqnarray}

We close this part by  mentioning the existence of a complementary
regularization scheme,
the so-called  `Kounterterm' approach
\cite{Olea:2006vd}.
There are some important differences in this case.
First, the `Kounterterm' approach is more naturally associated with a variational
principle where the extrinsic curvature $K_{ij}$
is kept fixed on the boundary.
Also, the boundary counterterms are constructed in terms of both the extrinsic and intrinsic 
curvature tensors.
The case of a CS-AdS$_5$ gravity is discussed in \cite{Mora:2004kb},
the corresponding `Kounterterm' expression being 
\begin{eqnarray} 
\label{Kt}
I_{\mathrm{Kt}} 
&=&
-\frac{1}{8\pi G}\frac{\ell^2}{8}
\int_{\partial \mathcal{M}} d^4x
\sqrt{-\gamma}
\bigg [ \delta^{[a_1a_2a_3]}_{ [b_1b_2b_3]}K_{a_1}^{b_1}
( 
\mathsf{R}_{a_2 a_3}^{b_2 b_3}
-K_{a_2}^{b_2}K_{a_3}^{b_3}
+\frac{2}{3\ell^2}\delta_{a_2}^{b_2}\delta_{a_3}^{b_3}
)
 \bigg ].
\end{eqnarray}
The action of the model consists in this case the sum of  
 (\ref{EGB0}) together with the boundary term (\ref{Kt}) 
 (note that the usual boundary term  (\ref{Ib1}) is absent  
 due to the choosen set of boundary conditions).
Given a solution of the field equations,
once the action is computed in this way,
the global charges can be computed using the Noether theorem, or 
by using the Euclidean black hole thermodynamics.
Further details on this formalism together with 
some examples and the generalization to the case $d=2n+1$ $(n\geq 2)$
can be found in 
\cite{Miskovic:2007mg},
\cite{Kofinas:2007ns}.

\section{Static black objects in  $d=5$ CS theory}

\subsection{The Schwarzschild-CS solution }
The counterparts of the Schwarzschild solution for a $d=5$
CS theory of gravity have been discussed already in the literature.
For completness, we review here their basic properties,
together with a derivation of their mass and entropy by using
the general formalism described above.   

Here one starts by considering the following line element
\begin{eqnarray}
\label{SGB}
ds^2=\frac{dr^2}{f(r)}+r^2 d\Sigma^2_{k,3}-b(r) dt^2
\end{eqnarray}%
where  $d\Sigma^2_{k,3}$ is the line element of a three-dimensional manifold $\Sigma_{k,3}$
\begin{equation}
d\Sigma^2_{k,3} =\left\{ \begin{array}{ll}
\vphantom{\sum_{i=1}^{3}}
 d\Omega^2_{3}& {\rm for}\; k = +1\\
\sum_{i=1}^{3} dx_i^2&{\rm for}\; k = 0 \\
\vphantom{\sum_{i=1}^{3}}
 d\Xi^2_{3} &{\rm for}\; k = -1\ ,
\end{array} \right.
\end{equation}
$d\Omega^2_{3}$ denoting the unit metric on $S^3$; by $H^3$ 
we will understand the three--dimensional hyperbolic space.
 
The  solution with 
\begin{eqnarray}
\label{back}
b(r)=f(r)=\frac{2r^2}{\ell^2}+k,
\end{eqnarray}%
corresponds to the AdS backgrounds in CS theory.
These configurations share all properties of their Einstein gravity  counterparts 
(for example they are also maximally symmetric).
Similar to that case \cite{Balasubramanian:1999re}, these backgrounds have a nonvanishing mass 
\begin{eqnarray}
\label{Mc}
 {\cal M}_c^{(k)}=-\frac{V_{k,3}}{8 \pi G}\frac{3k^2 \ell^2}{8},
\end{eqnarray}%
(with $V_{k,3}$ is the (dimensionless) volume associated with the   metric $d\Sigma^2_{k,3}$)
which, within the AdS/CFT correspondence, is interpreted as a Casimir term.

The Schwarzschild-CS black hole solutions have a very simple form which resembles the $d=3$
BTZ metric  \cite{Banados:1992wn} and have been studied
by many authors \cite{Cai:1998vy}, \cite{Crisostomo:2000bb}, \cite{Aiello:2004rz}.
Within the metric ansatz (\ref{SGB}), they are found for 
\begin{eqnarray}
\label{eq2n}
b(r)=f(r)=\frac{2 }{\ell^2}(r^2-r_H^2),
\end{eqnarray}%
with $r_H>0$ a 
constant corresponding to the event horizon radius.
A straightforward computation shows 
the absence of singularities for $r>r_H$. 
However, different from the Einstein gravity case,  
 the limit $r_H\to 0$ is singular for $k\neq 0$. 

The mass ${\cal M}$, temperature $T_H$ and entropy $S$ of this solution are
 \begin{eqnarray}
\label{prop-SGB}
{\cal M}=\frac{V_{k,3}}{8 \pi G}
\frac{3}{2}
r_H^2(k+ \frac{r_H^2}{\ell^2}),~~
T_H= \frac{r_H}{\pi \ell^2},
~~
S=\frac{V_{k,3}}{4 G}r_H^3(1+\frac{3k}{2}\frac{\ell^2}{r_H^2}).
\end{eqnarray}%
 Based on these relations, one can write the relatively simple equation of state
 (analogous to $f(p, V, T)$, for, say, a gas at pressure
$p$ and volume $V$)
  \begin{eqnarray}
\label{eq-state}
{\cal M}=\frac{3}{4}T_H S 
\left(
1-\frac{2k}{3}\frac{1}{1+k\sqrt{1+c_0 T_H S}}
\right),
\end{eqnarray}%
with $c_0=64 \pi G/(9 \ell^2V_{k,3})$. 
Note that, as usual, one can isolate the Casimir contribution to the total mass  ${\cal M}$
by writting 
 \begin{eqnarray}
{\cal M}=
{\cal M}_0^{(k)} +{\cal M}_c^{(k)},~~
{\rm where}~~
{\cal M}_0^{(k)}=\frac{3V_{k,3} }{8\pi G}\frac{\ell^2}{8}(k+\frac{2r_H^2}{\ell^2})^2.
\end{eqnarray}%

These black holes have some special properties and have been extensively studied in the literature.
For example, 
different from the Einstein gravity case, the $k=0,1$ solutions have a strictly positive specific heat,
since
 \begin{eqnarray}
\label{C-SGB}
C=T_H\frac{\partial S}{\partial T_H}=\frac{3 V_{k,3}}{4 G r_H}(1 +k \frac{\ell^2}{2r_H^2})
=\frac{3\pi  V_{k,3} \ell^4}{8 G}(k+2\pi^2\ell^2 T_H^2) T_H,
 \end{eqnarray}%
while the $k=-1$ black holes  become thermally unstable for small enough temperatures.
Note also that the $k=-1$
background is a special case of the general solution (\ref{eq2n}) 
with $r_H=\ell/\sqrt{2}$
and possesses a nonvanishing temperature and entropy.
  
It is easy to see that for these solutions,
the boundary metric upon which the dual field theory resides
corresponds to a  static Einstein universe in four dimensions, with a line element
$h_{ab}dx^adx^b= \frac{\ell^2}{2} d\Sigma^2_{k,3}-dt^2$.
The
stress tensor for the boundary dual theory, as computed according to (\ref{Tik-CFT}) has a vanishing trace, with  
  \begin{eqnarray}
\nonumber
8 \pi G  <\tau^{a}_b> =&&
U
\left( \begin{array}{cccc}
1&0&0&0
\\
0&1&0&0
\\
0&0&1&0
\\
0&0&0&-3
\end{array}
\right),
\end{eqnarray}
with $U=\frac{\sqrt{2}}{\ell^5} r_H^2(r_H^2+ k \ell^2)$
for black holes and $U=-\frac{k^2}{2\sqrt{2}\ell}$ for the AdS backgrounds (here $x^4=t)$.

 We close this part by remarking that, as discussed in \cite{Dotti:2007az},
 the solutions 
 (\ref{back}), (\ref{eq2n}) do not exhaust all possibilities
 allowed by the metric ansatz (\ref{SGB}). 
 For example, more complicated solutions describing
 wormholes and  'spacetime horns'  do also exist.
 Moreover, rather unusual, 
 due to a degeneration of the field equations,
the choice  $f(r)=\frac{2r^2}{\ell^2}+k$ provides
  a solution of the field equation
  for any expression of the redshift function $b(r)$.

\subsection{Squashed black hole solutions}
Interestingly, we have a found that the black hole 
solutions (\ref{SGB})
admit 'squashed' generalizations.
Although the topology of the horizon remains the same, 
the shape of the horizon is changed, as well as the boundary metric.

 Squashed  black hole solutions have been originally proposed in the context of
$d=4+1$ Kaluza-Klein theory with a vanishing cosmological constant.
Such configurations enjoyed considerable interest 
following the discovery  
of an exact solution in the five dimensional Einstein-Maxwell theory  \cite{Ishihara:2005dp}.
The horizon of a squashed black hole in  Kaluza-Klein theory has $S^3$ topology,  
while its spacelike infinity is a squashed sphere or a
$S^1$ bundle over $S^2$.  

Of interest here are squashed  black holes in AdS spacetime, 
considered in \cite{Murata:2009jt},  \cite{Brihaye:2009dm}
 within the Einstein gravity framework.
These solutions have a number of interesting properties; in particular they provide the gravity
dual for a ${\cal N} = 4$ super Yang-Mills theory on a background whose spatial part is a
squashed three sphere.
They are also relevant in connection to the so-called 'fragility' of the AdS black holes. 
As described in \cite{McInnes:2010ti} it turns out that AdS black
holes can become unstable to stringy effects when their horizon geometries are sufficiently distorted.

However, different from the case of a Kaluza-Klein theory \cite{Ishihara:2005dp},
no exact solutions could be constructed in the presence of a negative cosmological constant.
The configurations in \cite{Murata:2009jt},  \cite{Brihaye:2009dm}  
have been constructed numerically, by matching the
 near-horizon expansion of the metric to their asymptotic Fefferman-Graham
form \cite{Feff-Graham}.

The squashed black hole solutions in CS-AdS theory are found within the same metric ansatz as in 
the Einstein gravity case 
  \cite{Brihaye:2009dm}:
\begin{eqnarray}
\label{metric-squashed} 
ds^2= \frac{dr^2}{f(r)}+r^2 \big(d\theta^2+F_k^2(\theta)  d\varphi^2 \big)
+a(r)
\big(
dz+4n F_k^2 (\frac{\theta}{2} )d\varphi
\big )^2-b(r)dt^2,
\end{eqnarray}
with $F_{k}(\theta )$ a function which is fixed by the discrete parameter $k=0,\pm 1$:
\begin{eqnarray}  
\label{Fk}
F_{k}(\theta )=\left\{ 
\begin{array}{ll}
\sin \theta , & \mathrm{for}\ \ k=1 \\ 
\theta , & \mathrm{for}\ \ k=0 \\ 
\sinh \theta , & \mathrm{for}\ \ k=-1.%
\end{array}%
\right.
\end{eqnarray}%
As usual, $r$ and $t$ in the line element (\ref{metric-squashed}) are radial and time coordinates, respectively.
For $k=1$, $\theta$ and $\varphi$ are the spherical coordinates with the usual range $0\leq \theta \leq \pi $, $0\leq \varphi \leq 2 \pi$;
for $k=0$ and $k=-1$, the range of $\theta$ is not restricted, 
while $\varphi$ is still a periodic coordinate with $0\leq \varphi \leq 2 \pi$.
Then $d\Omega_{k,2}^2=d\theta^2+F_k^2(\theta) d\varphi^2$
is the metric on a two-dimensional surface of constant curvature $2k$;  
when $k = 0$, a constant $(r,t)$ slice is a flat surface,
while for $k = -1$, this sector is a space with constant negative curvature, also known as a hyperbolic
plane.

The periodicity $L$ of the coordinate $z$ is fixed for $k=1$
only,
in which case $z$
becomes essentially an Euler angle,
with $L=8\pi n$, as imposed by the absence of conical singularities.
Then a surface of constant $(t,r)$ is a squashed, topologically $S^3$ sphere.

Given this ansatz, we have found that the CS equations (\ref{eqs})
admit the following exact solution\footnote{We have found that 
 the metric ansatz (\ref{metric-squashed})
allows for other solutions than (\ref{ex-bh-backgr}), (\ref{ex-bh-sq}).
In fact, it seems that all configurations discussed in \cite{Dotti:2007az}
 possess squashed generalizations.
} 
\begin{eqnarray}  
\label{ex-bh-backgr}
 f(r)=b(r)=\frac{2r^2}{\ell^2}+\frac{k}{3}-\frac{2n^2}{3\ell^2},~~a(r)=\frac{2r^2}{\ell^2}~.
 \end{eqnarray}%
One can easily verify that the curvature invariants are finite everywhere, 
in particular at $r=0$.
Thus it is natural to interpret (\ref{ex-bh-backgr}) as the squashed deformation of the general AdS background (\ref{back}).
These solutions possess a nonvanishing  mass
\begin{eqnarray}
\label{Mc-sq}
 {\cal M}_c^{(k)}=-\frac{V_{k,2}\ell L}{ 96 \sqrt{2} \pi G}(\frac{2n^2}{\ell^2}-k)^2,
\end{eqnarray}%
(with $V_{k,2}$   the total area of the ($\theta,\varphi)$ surface)
which is again interpreted as a Casimir term.

The  squashed black holes have
\begin{eqnarray}  
\label{ex-bh-sq}
 f(r)=b(r)=\frac{2r^2}{\ell^2}-\frac{2r^2_H}{\ell^2}, ~~a(r)=\frac{2r^2}{\ell^2},
 \end{eqnarray}%
possesing an event horizon located at $r=r_H>0$.
 One can easily verify the absence of singularites for any $r\geq r_H$
 (although the limit $r_H \to 0$ is again pathological).
 
 The parameter $n$ is an input constant of the model, which is not fixed $apriori$.
 For $k=1$ a value of interest is 
 \begin{eqnarray}  
\label{val-n}
n=\frac{\ell}{2\sqrt{2}},
 \end{eqnarray}
  in which case 
 the surface of constant $r,t$ is a round sphere $S^3$
 and the Schwarschild-CS black hole (\ref{eq2n}) is recovered\footnote{Note that the normalization of the metric on $S^3$
 in (\ref{metric-squashed}) differs from that used in (\ref{SGB}). The radial and the time coordinates are also different.}.
This configuration separates prolate metrics from the oblate case ($n>\frac{\ell}{2\sqrt{2}}$). 
 
A straightforward computation based on the general formalism
in Section 2 leads to the following expressions
for the mass, temperature and entropy of 
these black holes\footnote{We have verified that similar results are obtained when 
using instead the  'Kounterterm' regularization procedure.
This holds also for the black string limit of the solutions in this work.}
\begin{eqnarray}  
\label{ex-bh-quant}
&&
{\cal M}=\frac{1}{8\pi G}\frac{V_{k,2}r_H^2 }{\sqrt{2} }\frac{L}{\ell}
\left( k-\frac{2n^2}{\ell^2}+\frac{3r_H^2}{\ell^2} \right),
\\
\nonumber
&&
T_H=\frac{r_H}{\pi \ell^2}, ~~S=\frac{r_H^3 V_{k,2}}{3\sqrt{2} G}\frac{L}{\ell}
\left(1+\frac{k\ell^2}{2r_H^2}-\frac{n^2}{r_H^2} \right),
 \end{eqnarray}%
 the first law of thermodynamics being satisfied.
 Again, one can isolate the Casimir term in the expression of the mass by writing
 \begin{eqnarray}  
\label{ex-bh-quant1}
{\cal M}={\cal M}_0^{(k)}+{\cal M}_c^{(k)},~~{\rm where}~~
{\cal M}_0^{(k)}=\frac{L\ell V_{k,2}}{96 \sqrt{2} \pi G} \left(\frac{6r_H^2}{\ell^2}+k-\frac{2n^2}{\ell^2} \right)^2,
\end{eqnarray}%
with ${\cal M}_c^{(k)}$ given by (\ref{Mc-sq}).

The  specific heat of the squashed black holes is
\begin{eqnarray}  
\label{ex-bh-Cv}
C=\frac{\pi \ell LV_{k,2}}{4\sqrt{2}G}T_H 
\left( 6\pi^2 \ell^4 T_H^2+k\ell^2-2n^2
\right )~,
\end{eqnarray}%
and does not possess a fixed sign.
Also, the above relation implies that squashing a black hole may render the configuration thermally unstable
(note also that, for any $k$, the entropy $S$ becomes negative
 for large enough values of $n$,
 which appear to be a signal of pathological behaviour).

The boundary metric upon which the dual field theory resides
corresponds to a  squashed static Einstein universe in four dimensions
  \begin{eqnarray}
h_{ab}dx^a dx^b= \frac{\ell^2}{2} \big(d\theta^2+F_k^2(\theta)  d\varphi^2 \big)
+ 
\big(
dz+4n F_k^2 (\frac{\theta}{2} )d\varphi
\big )^2- dt^2~.
  \end{eqnarray}
Also, the
stress tensor for the boundary dual theory is traceless
  (where $x^1=z,~x^2=\theta ,~x^3=\varphi ,~x^4=t$)
  \begin{eqnarray}
\label{tik-2}
8 \pi G  <\tau^{a}_b> =&&
U_1
\left( \begin{array}{cccc}
1&0&0&0
\\
0&1&0&0
\\
0&0&1&0
\\
0&0&0&-3
\end{array}
\right)
+
U_2
\left( \begin{array}{cccc}
1&0&4nF^2_k(\frac{\theta}{2})&0
\\
0&0&0&0
\\
0&0&0&0
\\
0&0&0&-1
\end{array}
\right)
,
\end{eqnarray}
  with $U_1=\frac{\sqrt{2}}{\ell^5} r_H^2(r_H^2+ 2 n^2)$, 
$U_2=-\frac{8n^2-k\ell^2}{2n^2+r_H^2}$
for squashed black holes and 
$U_1=\frac{(2n^2-k\ell^2)(14 n^2-k\ell^2)}{18\sqrt{2}\ell^5}$,
$U_2=-\frac{6(8n^2-k\ell^2)}{(14n^2-k\ell^2)}$
  for the squashed AdS backgrounds.


\subsection{Uniform black strings}
 Taking the limit $n=0$ in (\ref{metric-squashed}), (\ref{ex-bh-sq}),
we find the following exact solution 
\begin{eqnarray}
\label{UBS-BH}
ds^2=\frac{dr^2}{\frac{2}{\ell^2}(r^2-r_H^2)}
+r^2 \left( d\theta^2+F_k^2(\theta) d\varphi^2 \right)
+\frac{2r^2}{\ell^2} dz^2
-\frac{2}{\ell^2}(r^2-r_H^2) dt^2~,
\end{eqnarray}
describing an uniform AdS black string in $d=5$ CS gravity.
The properties of these solutions are rather special as compared to the squashed case.
For example, while the range of $\theta$ and $\phi$ is similar to the squashed black hole case,
for black strings with any $k$,
the  period $L$ of the compact 'extra'-direction $z$ is an arbitrary positive constant which plays no role in 
 our results\footnote{ However, similar to the $\Lambda=0$ Kaluza-Klein theory, the value of $L$ 
becomes relevant
 when discussing the issue of Gregory-Laflamme instability  of these objects \cite{Brihaye:2007ju}, 
 \cite{Delsate:2008kw}.}.

This kind of asymptotically locally-AdS solutions have been considered for the first time by
Horowitz and Copsey  in  \cite{Copsey:2006br},  for  configurations with $k=1$ only.
These black strings 
have an event horizon topology $S^{2}\times S^1$,
their conformal boundary being the product of 
time and $S^{2}\times S^1$.
These solutions have been generalized  in \cite{Mann:2006yi}
 to the case of an event horizon topology\footnote{The $k=0$ AdS black strings correspond  to planar black holes.}
$H^{2}\times S^1$, black strings in $d\geq5$ dimensions being considered as well. 
 AdS black strings with gauge fields are studied in 
\cite{Chamseddine:1999xk}, \cite{Brihaye:2007vm}, \cite{Brihaye:2007jua}, \cite{Bernamonti:2007bu}. 
The Ref. \cite{Brihaye:2007vm} discussed  also the properties of a special set of spinning black string solutions.
The issue of Gregory-Laflamme instability \cite{Gregory:1993vy}
for AdS black strings in Einstein  gravity
was addressed in \cite{Brihaye:2007ju}, 
nonuniform solutions ($i.e.$ with dependence on the compact 
`extra'-dimension $z$) being constructed in \cite{Delsate:2008kw}, \cite{Delsate:2009bd}.

As argued in \cite{Copsey:2006br}, \cite{Mann:2006yi}, such solutions in $d-$dimensions (with $d\geq 5)$
provide the gravity dual of a field theory  on
a $S^{d-3}\times S^1\times S^1$ (or $H^{d-3}\times S^1\times S^1$)
background.
Different from the $\Lambda=0$ limit, it was found in 
\cite{Copsey:2006br},
\cite{ Mann:2006yi} that the AdS black 
string solutions  with an event horizon 
topology $S^{d-3}\times S^1$ have a nontrivial, vortex-like
globally regular limit with zero event 
horizon radius.

However, a general feature of all these AdS black strings/vortices  is the absence of exact solutions 
(see, however, the extremal Einstein-Maxwell configurations 
in \cite{Chamseddine:1999xk}).
Remarkably, as shown by (\ref{UBS-BH}),
the extension of the gravity to a CS model allows for
exact solutions.

Let us start by discussing first another
set of solutions, which provide a natural background
for the black strings (\ref{UBS-BH}).
The corresponding expression of the line element, as resulting 
from (\ref{ex-bh-backgr}), reads
\begin{eqnarray}
\label{UBS-soliton}
ds^2=\frac{dr^2}{\frac{2r}{\ell^2} +\frac{k}{3}}
+r^2 \left( d\theta^2+F_k^2(\theta) d\varphi^2 \right)
+\frac{2r^2}{\ell^2} dz^2
-(\frac{2r}{\ell^2} +\frac{k}{3}) dt^2.
\end{eqnarray}
One can see that the $k=1$ solution describe 
the CS counterparts of the Einstein-gravity vortices in \cite{Mann:2006yi};
the $k=0$ solution is already contained in (\ref{SGB}), (\ref{back}),
while the $k=-1$ case corresponds to a special black string in (\ref{UBS-BH}).

The computation of the boundary stress tensor $T_{ab}$ based on the relations in Section 2 is again 
straightforward.    
Apart from the mass-energy ${\cal M}$, the solutions possess
this time a second 
 charge associated with the compact $z$ direction, corresponding 
 to a tension ${\mathcal T}$. 
Then we find the following expressions for the  mass and tension of the background (\ref{UBS-soliton})
\begin{eqnarray}
\label{UBS-soliton-MT}
{\cal M}_c^{(k)}=-\frac{L\ell V_{k,2}}{96\pi G\sqrt{2}}k^2,~~
{\cal T}_c^{(k)}=-\frac{5\ell V_{k,2}}{288\pi G\sqrt{2}}k^2.
\end{eqnarray}
Returning to the general black string solution (\ref{UBS-BH}),
the corresponding expressions for  conserved global charges read
 \begin{eqnarray}
{\cal M}=\frac{V_{k,2}L}{8 \pi G}\frac{r_H^2(3 r_H^2+k \ell^2)}{\sqrt{2}\ell^3},~~
{\cal T}=\frac{V_{k,2}}{8 \pi G}\frac{r_H^2(  r_H^2+k \ell^2)}{\sqrt{2}\ell^3},
\end{eqnarray}
which can also be written as
 \begin{eqnarray}
 &&
{\cal M}={\cal M}_0^{(k)}+{\cal M}_c^{(k)},~~{\rm where}~~
{\cal M}_0^{(k)}=\frac{L\ell V_{k,2}}{96\sqrt{2}\pi G}\left(\frac{6r_H^2}{\ell^2}+k \right),
\\
\nonumber
 &&
{\cal T}={\cal T}_0^{(k)}+{\cal T}_c^{(k)},~~{\rm where}~~
{\cal T}_0^{(k)}=\frac{L\ell V_{k,2}}{288\sqrt{2}\pi G}
\left(\frac{6r_H^2}{\ell^2}+k \right)
\left(\frac{6r_H^2}{\ell^2}+5 k \right),
\end{eqnarray}
with ${\cal M}_c^{(k)}$, ${\cal T}_0^{(k)}$ given by (\ref{UBS-soliton-MT}).

The Hawking temperature and entropy of these solutions are given by
\begin{eqnarray}
T_H=\frac{r_H}{\pi \ell^2},~~S=\frac{V_{k,2}L}{4\sqrt{2}G \ell}r_H(2r_H^2+k\ell^2).
\end{eqnarray}
Similar to the Einstein gravity case \cite{Mann:2006yi}, the  black strings satisfy also a simple 
Smarr-type formula, relating quantities defined at 
infinity to quantities defined at the event horizon:
\begin{eqnarray}
\label{smarrform} 
{\cal M}+{\mathcal T}L=T_H S~,
\end{eqnarray} 
the first law of thermodynamics
\begin{eqnarray}
\label{1st-law-UBS} 
 d{\cal M}=T_H dS-{\cal T}dL~,
\end{eqnarray}
being also fulfilled.
Also, one can easily see that the thermodynamics of these solutions is rather similar to the (un-squashed) black hole case 
(for example the $k=1$ solutions have a positive specific heat).

Let us also mention that the boundary metric upon which the dual field theory resides
is
$
h_{ab}dx^a dx^b= \frac{\ell^2}{2} \big(d\theta^2+F_k^2(\theta)  d\varphi^2 \big)
+ 
dz ^2- dt^2.
$
The  expression of the
stress tensor for the boundary dual theory
is be  found by taking the limit $n=0$ in (\ref{tik-2}).

\section{Rotating black holes with spherical horizon topology 
and two equal angular momenta}
On physical grounds, it is natural to expect the existence of
rotating generalizations of the static black holes discussed in Section 3.1.
However, the presence of the GB term in the action makes highly non-trivial 
the task of finding a closed form of these solutions\footnote{Some results in this direction 
are reported in \cite{Anabalon:2009kq}.
However, the solution there has rather special properties (for example the line element is not circular) 
and does 
not describe a black object.
}.

While rotating black holes will generically possess two independent
angular momenta and a more general topology 
of the event horizon,
restrincting the study to the special 
case of configurations with equal-magnitude angular momenta and a spherical horizon topology
leads to a  dramatic simplification of the problem.
As first observed in \cite{Kunz:2005nm},
these configurations have a symmetry enhancement 
which allows to factorize the angular dependence.
This results in a system of ordinary differential equations
which can be easily studied.

The suitable metric ansatz in this case is similar to that used 
in the previous work  \cite{Brihaye:2008kh}, \cite{Brihaye:2010wx}, with
\begin{eqnarray}
\label{metric}
&&ds^2 = \frac{dr^2}{f(r)}
  + g(r) d\theta^2
+h(r)\sin^2\theta \left( d \varphi_1 -w(r)dt \right)^2 
+h(r)\cos^2\theta \left( d \varphi_2 -w(r)dt \right)^2 ~~{~~~~~}
\\
\nonumber
&&{~~~~~~}+(g(r)-h(r))\sin^2\theta \cos^2\theta(d \varphi_1 -d \varphi_2)^2
-b(r) dt^2,
\end{eqnarray}
where $\theta  \in [0,\pi/2]$, $(\varphi_1,\varphi_2) \in [0,2\pi]$, 
and $r$ and $t$ denote the radial and time coordinate, respectively.
Note that this metric ansatz admits a simpler expression in terms of
the left-invariant
1-forms $\sigma_i$ on $S^3$, with 
\begin{eqnarray}
\label{metric2}
&&ds^2 = \frac{dr^2}{f(r)}
  + \frac{1}{4}g(r)(\sigma_1^2+\sigma_2^2)+\frac{1}{4}h(r) \big(\sigma_3-2w(r) dt \big)^2
-b(r) dt^2,
\end{eqnarray}
where 
$\sigma_1=\cos \psi d\bar \theta+\sin\psi \sin  \bar\theta d \phi$, 
$\sigma_2=-\cos \psi d\bar \theta+\cos\psi \sin \bar \theta d \phi$,
$\sigma_3=d\psi  + \cos \bar \theta d \phi$ 
and $2\theta=\bar \theta$,
 $\varphi_2-\varphi_1=\phi$,
  $\varphi_1+\varphi_2=\psi$.
For such solutions the isometry group is enhanced from $R \times U(1)^{2}$
to $R \times U(2)$, where $R$ denotes the time translation\footnote{The Myers-Perry-AdS (MP-AdS) solution  \cite{Hawking:1998kw}
with two angular momenta can also be written in this form, with
$f(r)=1+r^2/\ell^2-2M\Sigma/r^2+2Ma^2/r^4$,
$h(r)=r^2(1+2Ma^2/r^4)$,
$w(r)=2Ma/(r^2h(r))$,
$b(r)=r^2f(r)/h(r)$
and $g(r)=r^2$.
Here $M$ and $a$ are two constants while $\Sigma=1-a^2/\ell^2$.
 }.

In the following we fix the metric gauge by taking $g(r)=r^2$.
Then the field equations (\ref{eqs})
reduce to a  set of four ordinary differential
equations for the functions $b,f,h$ and $w$ plus one extra constraint equation.
The explicit form of these equations is given in Appendix A, together with a discussion of the 
issue of constraint.

We close this part by giving  some useful relations which 
follow directly from the eqs. (\ref{eq1})-(\ref{eq5}) and which are relevant for
 computing  the action of solutions: 
\begin{eqnarray}
\label{totder1} 
&&\frac{1}{\sin^2 \theta}(R_{\varphi_1}^t+\frac{\ell^2}{8}H_{\varphi_1}^t)
=\frac{1}{\cos^2 \theta}(R_{\varphi_2}^t+\frac{\ell^2}{8}H_{\varphi_2}^t)
\\
\nonumber
&&
{~~~~~~~~~~~~~~~~~~~~~~~~~}=  \frac{1}{2 r^2} \sqrt{\frac{f}{ b h}}\frac{d }{dr}  
 \left [\sqrt{\frac{f h}{b}} h w'
 \bigg (
- r^2 +  \frac{\ell^2}{2} ( f-4+ \frac{3h}{r^2})\bigg)
\right],
\\ 
\label{totder2} 
&&R_t^t+\frac{\ell^2}{8}(H_t^t+\frac{1}{2}L_{GB})=
\frac{1}{2r^2} \sqrt{\frac{f}{ b h}}  \frac{d }{dr}
 \bigg [\sqrt{\frac{f h}{b}}
 \bigg( r^2(hww'-b')
 \\
 \nonumber
&&
{~~~~~~~~~~~~~~~~~~~~~~~~~~~~~~~}
+\frac{\ell^2}{2}\left((f-4+\frac{h}{r^2}+\frac{rfh'}{h})b'
 +(4-\frac{3h}{r^2}-f)hww'+rfhw'^2\right)
  \bigg )
\bigg ] ,
\end{eqnarray} 
where 
$H_{\mu \nu}=2(R_{\mu \sigma \kappa \tau }R_{\nu }^{\phantom{\nu}%
\sigma \kappa \tau }-2R_{\mu \rho \nu \sigma }R^{\rho \sigma }-2R_{\mu
\sigma }R_{\phantom{\sigma}\nu }^{\sigma }+RR_{\mu \nu })-\frac{1}{2}%
L_{GB}g_{\mu \nu } 
$
is the Lanczos (or the Gauss-Bonnet) tensor.
\subsection{Asymptotic form of the solutions}

The horizon of these spinning black hole resides at the constant value of
the radial coordinate $r=r_H>0$, and is characterized by the condition
\begin{eqnarray}
\label{c0}
f(r_H)=b(r_H) =  0. 
\end{eqnarray} 
Restricting to nonextremal solutions, the following expansion
holds near the event horizon: 
\begin{eqnarray}
\label{c1}
&&f(r)=f_1(r-r_H)+  O(r-r_H)^2,~~h(r)=h_H+ O(r-r_H),
\\
\nonumber
&&
b(r)=b_1(r-r_H)+O(r-r_H)^2,~~w(r)=\Omega_H+w_1(r-r_H)+O(r-r_H)^2.~{~ }
\end{eqnarray}  
For a given event horizon radius, 
the essential parameters characterizing the event horizon
are $f_1,~b_1$, $\Omega_H$ and $w_1$ (with $f_1>0,~b_1>0$),
which fix all other coefficients in (\ref{c1}), including those of the higher order terms not displayed there.
The explicit form of these   coefficients is very complicated,
except for  $h_H$ which has the  
expression\footnote{Note the unusual dependence of $h(r_H)$ on $w'(r_H)$.
The corresponding expression for the Einstein gravity solution reads $h_H=r_H^2(4-f_1 r_H+4r_H^2/\ell^2)/2$.}
\begin{eqnarray}
\label{h_H}
h_H=\sqrt{\frac{b_1}{f_1}} \frac{r_H}{\ell^2 w_1}(4r_H-f_1\ell^2).
 \end{eqnarray} 
 The metric of a spatial cross section of the horizon is 
\begin{eqnarray}
\label{metric-eh}
ds^2_H= 
 \frac{1}{4}r_H^2 \big(  \sigma_1^2+\sigma_2^2 +\frac{h_H}{r_H^2}  \sigma_3^2 \big),
\end{eqnarray} 
 which corresponds to a squashed $S^3$ sphere. 
 
%
The (constant) horizon angular velocity $\Omega_H$
is defined in terms of the Killing vector
$\chi=\partial/\partial_t+
 \Omega_1 \partial/\partial \varphi_1 + \Omega_2 \partial/\partial \varphi_2 $
which is null at the horizon.
For the solutions within the ansatz (\ref{metric}), the 
horizon angular velocities are equal, $\Omega_1=\Omega_2=\Omega_H$.

One can also write 
an asymptotic form of the solutions, 
involving three free parameters ${\cal U},~{\cal V}$ and ${\cal W}$,
with
\begin{eqnarray}
\nonumber
&&
f(r) = \frac{2 r^2}{\ell^2} + 1 +{\cal V}  
+f_2^{(as)}\frac{\ell^2}{2r^2}+O(1/r^4),
~~b(r) = \frac{2 r^2}{\ell^2} + 1 + {\cal U} +
b_2^{(as)}\frac{\ell^2}{2r^2}+O(1/r^4),
\\
&&
\label{BH-rot-inf}
h(r) = r^2( 1  + h_2^{(as)} \frac{2\ell^2}{r^2}) + O(1/r^2),~~
w(r) = {\cal W}\frac{ \ell^2}{2 r^2}+\frac{ \ell^4}{4 r^4}w_4^{(as)}+O(1/r^6).
\end{eqnarray}  
 All coefficients in these series (including the higher order ones)
 can be expressed as combinations of  ${\cal U},~{\cal V}$ and ${\cal W}$.
 One finds $e.g.$ 
\begin{eqnarray}
\nonumber
&&f_2^{(as)}=\frac{-18({\cal U}-{\cal V})^3 {\cal V}+(2{\cal U}^2+3 {\cal U}{\cal V}-6 {\cal V}^2)
\ell^2 {\cal W}^2 -\ell^4 {\cal W}^4}{(2{\cal U}-3{\cal V})(2{\cal U}({\cal U}-{\cal V})-\ell^2 {\cal W}^2},
\\
&&
\nonumber
b_2^{(as)}= \frac{1}{8}
\bigg(
2{\cal U}(6+{\cal U})-2({\cal U}+7){\cal V}-\frac{(4+3{\cal V})}{{\cal V}}\ell^2 {\cal W}^2
-\frac{4(2{\cal U}-3{\cal V})^2({\cal U}-{\cal V})}{2{\cal U}({\cal V}-{\cal U})+\ell^2 {\cal W}^2}
\bigg),
\\
\label{asympt-coeff}
&&
h_2^{(as)}=\frac{3({\cal U}-{\cal V}){\cal V}-\ell^2 {\cal W}^2}{2{\cal U}-3{\cal V}},
\\
&&
\nonumber
w_4^{(as)}=\frac{1}{4}\frac{{\cal W}}{2{\cal U}-3{\cal V}}
\bigg (
2(({\cal U}-4){\cal U}+3{\cal V}-7{\cal U}{\cal V}+6{\cal V}^2)
+\frac{(4+3 {\cal V})}{{\cal V}}\ell^2 {\cal W}^2
\\
\nonumber
&&
{~~~~~~~~~~~~~~~~~~~~~}
-\frac{4(2U-3V)^2(U-V)}{2U(V-U)+\ell^2 W^2}
\bigg),
\end{eqnarray}
the expression of other terms being too complicated to display it here.
Also, 
one can see that for these asymptotics, the boundary metric is ${\it not}$ rotating.
  
\subsection{Physical quantities}
 The  Hawking temperature  of these configurations is given by
\begin{eqnarray} 
\label{Temp-rot} 
  T_H=\frac{1}{4\pi}\sqrt{b'(r_H)f'(r_H)}~.
\end{eqnarray} 
The conserved 
charges  of the rotating black holes are obtained by 
using  again the counterterm  method in conjunction 
with the quasilocal formalism, as described in Section 2.  
These are given by the following complicated expressions
\begin{eqnarray}
&&
\label{mass-rot} 
{\cal M}={\cal M}_0+{\cal M}_{c}^{(1)},~~
{\rm with}~~{\cal M}_0=\frac{V_{1,3}}{8 \pi G}
\bigg [
\frac{\ell^2 {\cal V} }{8(2{\cal U}-3{\cal V})}
(3{\cal V}({\cal V}-2{\cal U})+4\ell^2 {\cal W}^2)
\bigg ],
\\
&&
\label{J-rot} 
J_1=J_2=\frac{V_{1,3}}{64 \pi G}(2-{\cal V})\ell^4 {\cal W},
\end{eqnarray}
where $V_{1,3}=2\pi^2$ and ${\cal M}_{c}^{(1)}$ given by (\ref{Mc}).
Thus in constrast to the case of spinning solutions in
Einstein gravity \cite{Hawking:1998kw} or in EGB model with $\alpha \neq \ell^2/2$ \cite{Brihaye:2008kh}, 
the constant ${\cal W}$ 
which enters the asymptotics of the metric functions $w(r)$ associated with rotation
($w(r)={\cal W}\ell^2/(2r^2)+\dots$)
does not fix alone the angular momentum.
Moreover,  ${\cal W}$ enters the expression of mass.

The computation of solutions' action is standard.
The bulk action evaluated on the equations of motion
can easily be computed
by replacing the $R+\frac{12}{\ell^2} +\frac{\ell^2}{8}L_{GB}$ volume term with
$2(R_t^t+\frac{\ell^2}{8}( H_t^t+\frac{1}{2}L_{GB} ))$.
Then one makes use of (\ref{totder2}) to express
the volume integral of this quantity as the difference 
of two boundary integrals.
The boundary integral on the event horizon is simplified by using the identity
(\ref{totder1}).
Then the divergencies of the boundary integral at infinity,  
together with the contribution 
from (\ref{Ib1}) are regularized by 
the counterterm (\ref{Lagrangianct}). 
Upon application of the Gibbs-Duhem relation to the partition 
function,  one finds  the entropy 
\begin{eqnarray}
\label{ent-rot} 
S=\beta ({\cal M}- 2\Omega_H J)-I_{cl}=\frac{V_{1,3}}{4G}r_H^2\sqrt{h_H}
\left( 
1+\frac{\ell^2}{2r_H^2}(4-\frac{h_H}{r_H^2})
\right).
\end{eqnarray}

With these quantities, the solutions
should satisfy the first law of thermodynamics (\ref{1stlaw}),
which, for this particular case reads
\begin{eqnarray}
\label{1stlaw-rot}
d{\cal M}=T_H dS+2\Omega_H dJ.
\end{eqnarray}

For completness, let us mention that
boundary metric upon which the dual field theory resides
corresponds to a  static Einstein universe with a line element
 $h_{ab}dx^a dx^b= \frac{\ell^2}{2} (d\theta^2+\sin^2 \theta d\varphi_1^2+ \cos^2 \theta d\varphi_2^2)-dt^2$.
The stress tensor for the boundary dual theory is again traceless with a rather complicated expression
(here $x^1=\theta$, $x^2=\varphi_1$, $x^3=\varphi_3$ and $x^4=t$):
\begin{eqnarray}
\nonumber
8 \pi G  <\tau^{a}_b> =&&
U_1
\left( \begin{array}{cccc}
1&0&0&0
\\
0&1&0&0
\\
0&0&1&0
\\
0&0&0&-3
\end{array}
\right)
+U_2 
\left( \begin{array}{cccc}
0&0&0&0
\\
0&\sin^2 \theta &\cos^2 \theta &0
\\
0&\sin^2 \theta &\cos^2 \theta &0
\\
0&0&0&-1
\end{array}
\right)
\\
\label{st1}
&&+
U_3
 \left( \begin{array}{cccc}
0&0&0&0
\\
0&0&0&0
\\
0&0&0&0
\\
0&\sin^2 \theta &\cos^2 \theta&0
\end{array}
\right)
+
U_4
 \left( \begin{array}{cccc}
0&0&0&0
\\
0&0&0&1
\\
0&0&0&1
\\
0&0&0&0
\end{array}
\right)
,
\end{eqnarray}
where
\begin{eqnarray}
\nonumber
&&
U_1=\frac{-4{\cal U}^2 {\cal V}+{\cal U}(6{\cal V}^2-2)+{\cal V}(3-3{\cal V}^2+2\ell^2 {\cal W}^2)}{2\sqrt{2}\ell(2 {\cal U}-3 {\cal V})},
~~U_2=\frac{-6({\cal U}-{\cal V})^2{\cal V}+\ell^2 {\cal V} {\cal W}^2}{\sqrt{2}\ell(-2{\cal U}+3{\cal V})},
\\
\label{Ui}
&&
U_3=-\frac{\ell}{\sqrt{2}}({\cal V}-2){\cal W},~~U_4=\frac{\sqrt{2}}{\ell}(2+{\cal V}) {\cal W}.
\end{eqnarray}

\subsection{Slowly rotating solutions}
We  strongly suspect that, in contrast to  the generic EGB case, the spinning
black hole solutions could be constructed in closed form
 for the CS model.
However, despite our effort,  we could not find this solution so far
(a spinning, singular configuration is reported in Appendix C).

Nevertheless, some progress in this direction can be achieved in the slowly rotating 
limit. Such solutions can be found by considering perturbation theory 
around the static solution (\ref{SGB}) in terms of a
small parameter $a$
which corresponds to the 
constant  ${\cal W}$   entering
 the asymptotic expansion (\ref{BH-rot-inf}).
 The second parameter of the solution is the horizon radius $r_H$.

Our results up to order seven in perturbation theory  
lead to conjecture that
the spinning black solutions admit an expression 
on the form\footnote{This corresponds to the most general (perturbative) solution of the equations of motion
compatible with both the near horizon expansion (\ref{c1}) and the asymptotic form (\ref{BH-rot-inf}).
However, a more general solution can be written when relaxing these assumptions.}
\begin{eqnarray}
\nonumber
&&
b(r)=b_0(r) 
\left(1+\sum_{k \geq 1}\frac{(-1)^{kP}a^k}{\beta^{3k}(1+2\beta^2)^{k-1}}\sum_{j=1}^k b_{kj}\beta^{2j}\left( \frac{\ell}{r}\right)^{2j} \right),
\\
\label{sol-slow}
&&
f(r)=f_0(r) 
\left(1+\sum_{k \geq 1}\frac{(-1)^{kP}a^k}{\beta^{3k}(1+2\beta^2)^{k-1}}\sum_{j=1}^k f_{kj}\beta^{2j}\left( \frac{\ell}{r}\right)^{2j} \right),
\\
\nonumber
&&
h(r)=h_0(r) 
\left(1+\sum_{k \geq 1}\frac{(-1)^{kP}a^k}{\beta^{3k}(1+2\beta^2)^{k-1}}\sum_{j=1}^k h_{kj}\beta^{2j}\left( \frac{\ell}{r}\right)^{2j} \right), 
\\
\nonumber
&&
w(r)= \frac{\beta}{\ell}
 \sum_{k \geq 1}\frac{(-1)^{(k+1)P}a^k}{\beta^{3k}(1+2\beta^2)^{k-1}}\sum_{j=1}^k w_{kj}\beta^{2j}\left( \frac{\ell}{r}\right)^{2j} , 
\end{eqnarray}
where
\begin{eqnarray}
\label{b-cvf}
b_0(r)=f_0(r)=\frac{2}{\ell^2}(r^2-r_H^2),~~h_0(r)=r^2,
 \end{eqnarray}
 are the metric functions of the static black hole solution and
 \begin{eqnarray}
\label{a-W}
a= \frac{1}{2}\ell {\cal W} 
 \end{eqnarray}
 is the expansion parameter.
Also, to simplify the relations we define the dimensionless parameter
  \begin{eqnarray}
\label{def-beta}
\beta=\frac{r_H}{\ell}.
 \end{eqnarray}
The parameters $b_{kj}$, $f_{kj}$, $h_{kj}$ and $w_{kj}$ in the expression (\ref{sol-slow})
are polynomials in $\beta^2$; their explicit form up to order four
is given in Appendix C.
Here we give only the first order expression of the solution\footnote{The corresponging expression in a general EGB theory  with $ \ell^2>2\alpha$  
has $b_1(r)=f_1(r)=h_1(r)=0$ and 
  \begin{eqnarray}
\label{a-W-MP}
\nonumber
w_1(r)=\frac{\ell^3}{2r^4}
\left (
\sqrt{1-\frac{2\alpha}{\ell^2}}
+\sqrt{1-\frac{2\alpha}{\ell^2}+\frac{\alpha}{r^4}(2\beta^2(1+\beta^2)\ell^2+\alpha)}
\right)^{-1}~,
  \end{eqnarray}
such that $w_1(r)\sim \frac{\ell^4}{4 r^4\sqrt{\ell^2-2\alpha}}$ asympotically (note the different power of $r$
as compared to  (\ref{BH-rot-inf})). 
However, in this case we could not construct a general perturbative solution similar to (\ref{sol-slow}).
}:
\begin{eqnarray}
&&
b(r)= \frac{2}{\ell^2} \big(r^2-\beta^2 \ell^2 \big),~~
~~
f(r)= \frac{2}{\ell^2} \big(r^2-\beta^2 \ell^2 \big)\bigg(1-(-1)^P \frac{a \ell^2}{2 \beta r^2} \bigg),
\label{sol-slow-1st}
\\
\nonumber
&&
h(r)=  r^2 \big(1-(-1)^P \frac{3 a \ell^2}{2 \beta r^2} \big),~
~~w(r)=\frac{a \ell}{r^2}. 
\end{eqnarray}
The relations (\ref{sol-slow}), (\ref{sol-slow-1st}) contain  an extra-parameter $P$, with $P=0$ or $P=1$,
which shows the existence of
two different  families of  solutions.
The difference between these two solutions becomes more transparent when writing
the expression of some relevant global quantities.
After replacing the expressions  (\ref{sol-slow}) in (\ref{Temp-rot}), (\ref{mass-rot}),  (\ref{J-rot}), (\ref{ent-rot})
one finds the general expressions\footnote{We have verified that these quantities satisfy the first law of thermodynamics
 up to order seven. Also the horizon remain regular
 up to that order.}:
\begin{eqnarray}
\nonumber
&&
{\cal M}={\cal M}^{(0)}+\frac{\pi \ell^2}{G}\sum_{k\geq 1}\frac{(-1)^{kP}a^k}{\beta^{3k-2}(1+2\beta^2)^{k-2}}M_k(\beta),
~~
J=\frac{\pi \ell^3}{16 G} \sum_{k \geq 1}\frac{(-1)^{(k+1)P}a^k}{\beta^{3k-3}(1+2\beta^2)^{k-1}}j_k(\beta),
\\
\nonumber
&&
\Omega_H=\frac{1}{\ell} \sum_{k \geq 1}\frac{(-1)^{(k+1)P} a^k}{\beta^{3k-1}(1+2\beta^2)^{k-1}}w_k(\beta),
~~
S=S^{(0)}+\frac{\pi^2 \ell^3}{16 G} \sum_{k \geq 1}\frac{(-1)^{kP}a^k}{\beta^{3k-1}(1+2\beta^2)^{k-2}}s_k(\beta),
\\
&&
\label{pt1}
T_H=T_H^{(0)}+\frac{1}{4\pi \ell} \sum_{k \geq 1}\frac{(-1)^{kP}a^k}{\beta^{3k-1}(1+2\beta^2)^{k-1}}t_k(\beta),
\end{eqnarray}
where 
${\cal M}^{(0)}$,  $S^{(0)}$ and  $T_H^{(0)}$ are the mass, entropy and temperature of the static black hole.

Here we give again only the first order expressions (these quantities up to order four are given in Appendix B; 
we only mention that $M_k$, $j_k$, $s_k$, $w_k$ and $t_k$ are polynomials
in $\beta^2$):
\begin{eqnarray}
\label{1st-order-expr}
\nonumber
&&
 {\cal M}= \frac{2\pi }{8G}\ell^2 \beta^2(1+\beta^2)-(-1)^P a\frac{\pi \ell^2}{G}\frac{3(1+2\beta^2)}{16\beta},~~~
J= a \frac{\pi \ell^3}{16G}(3+2 \beta^2),
\\
\nonumber
&&
\Omega_H= a\frac{1}{\ell \beta^2},~
~~
S= \frac{\beta(3+2\beta^2)\pi^2\ell^3}{4G}-(-1)^P a\frac{\pi^2\ell^3}{16 G}\frac{3(1+2\beta^2)}{\beta^2},~
~~~
T_H= \frac{\beta}{\pi \ell}-(-1)^P\frac{1}{\pi \ell}\frac{1}{4\beta^2}~.
\end{eqnarray}
Note that, different from the case of Myers-Perry solution, all physical relevant quantities receive corrections 
already in the first order of perturbation theory.
On a technical level, this can be attributed to the special form 
of the perturbative solution (\ref{sol-slow-1st}) and is anticipated already by the expressions (\ref{h_H}), (\ref{J-rot}).

Supposing $a>0$, one can see that for $P=0$
the mass, entropy and temperature of solutions decrease with $a$ (at least to lowest order).
Rather unusual, these quantities increase with $a$ for the second solution ($P=1$).
When changing the sign of $a$, the two solutions interchange.

\subsection{Nonperturbative solutions: numerical  results}

The non-perturbative solutions are constructed numerically,
by integrating the system of coupled non-linear 
ordinary differential equations given in Appendix A,
with appropriate boundary conditions  which follow from (\ref{c1}), (\ref{BH-rot-inf}).

To simplify the problem,
we restrict our integration to the region outside the event horizon, $r\geq r_H$.
In our approach, we use a 
 standard solver \cite{COLSYS},
 which involves a Newton-Raphson 
  method for 
boundary-value ordinary differential equations, 
equipped with an adaptive mesh selection procedure.
Typical mesh sizes include around 400 points.
Also, the solutions in this work have a typical relative accuracy of $10^{-6}$.

Somehow unexpected, we have found the numerical study of the spinning solutions in the 
$d=5$ CS theory more complicated than in the case of rotating solutions
of the generic EGB  model in 
\cite{Brihaye:2008kh}, 
\cite{Brihaye:2008xu},
\cite{Brihaye:2010wx}.
For example, for the solutions in this work we did not manage to fix the 
total angular momentum as an input parameter 
and thus we could not study in a systematic way
spinning black holes in a canonical ensemble (this can be attributed to the complicated expression (\ref{J-rot})
of $J$, with a dependence on both ${\cal V}$ and ${\cal W}$). 
Therefore in our numerical scheme,
for a given value of $\ell$,
 we have choosen $r_H$ and $\Omega_H$
as input parameters.
The numerical output provides the profiles of the functions $b,f,h$ and $w$
and their first derivative on some mesh.
By interpolating these profiles, we could compute all quantities of interest.

Our numerical results confirm the existence of spinning
generalizations of the spherically symmetric
black hole solutions also at the non-perturbative level\footnote{We have found that the perturbative solutions
provide a good approximation for a large part of the numerical configurations.}. 
 The profiles of a typical solution are presented of Figure 1 as a function of the radial coordinate.

  \newpage
\begin{figure}[ht]
\hbox to\linewidth{\hss%
	\resizebox{8cm}{6cm}{\includegraphics{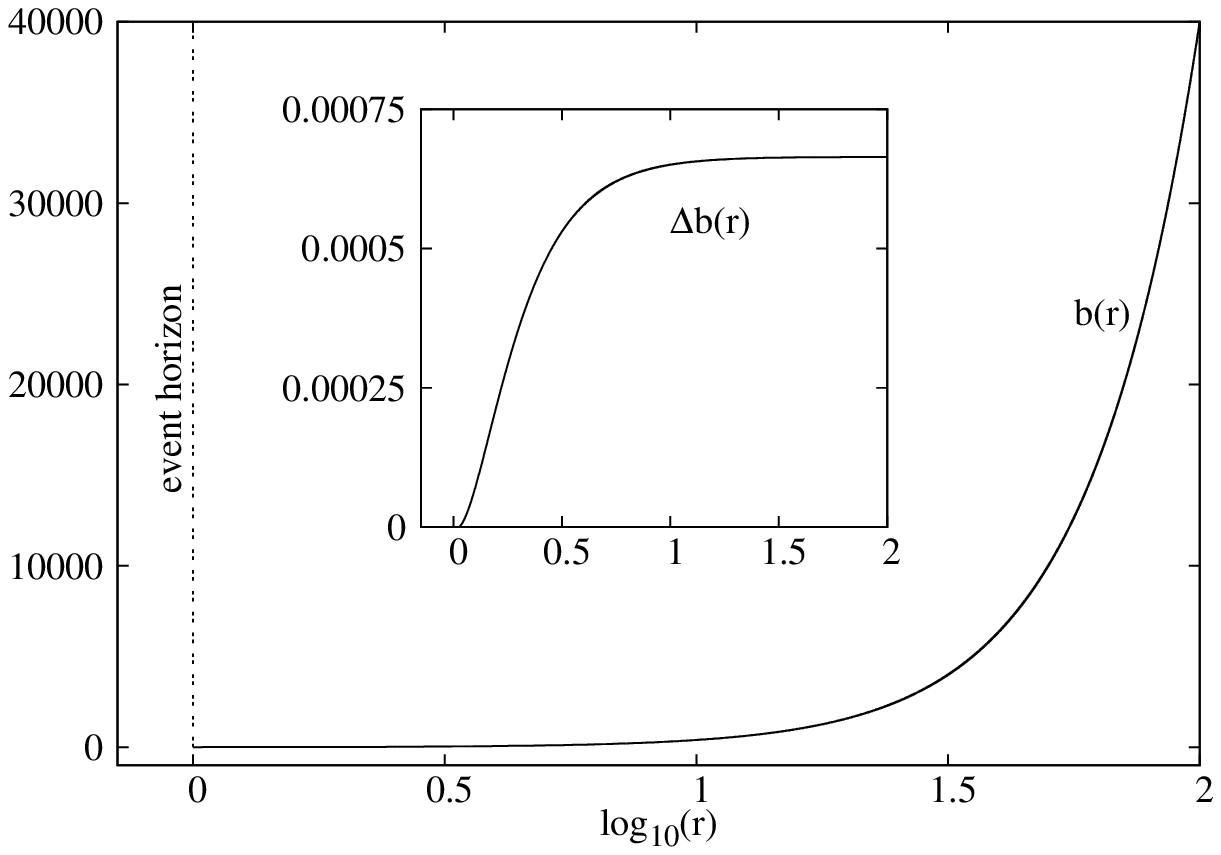}}
\hspace{10mm}%
        \resizebox{8cm}{6cm}{\includegraphics{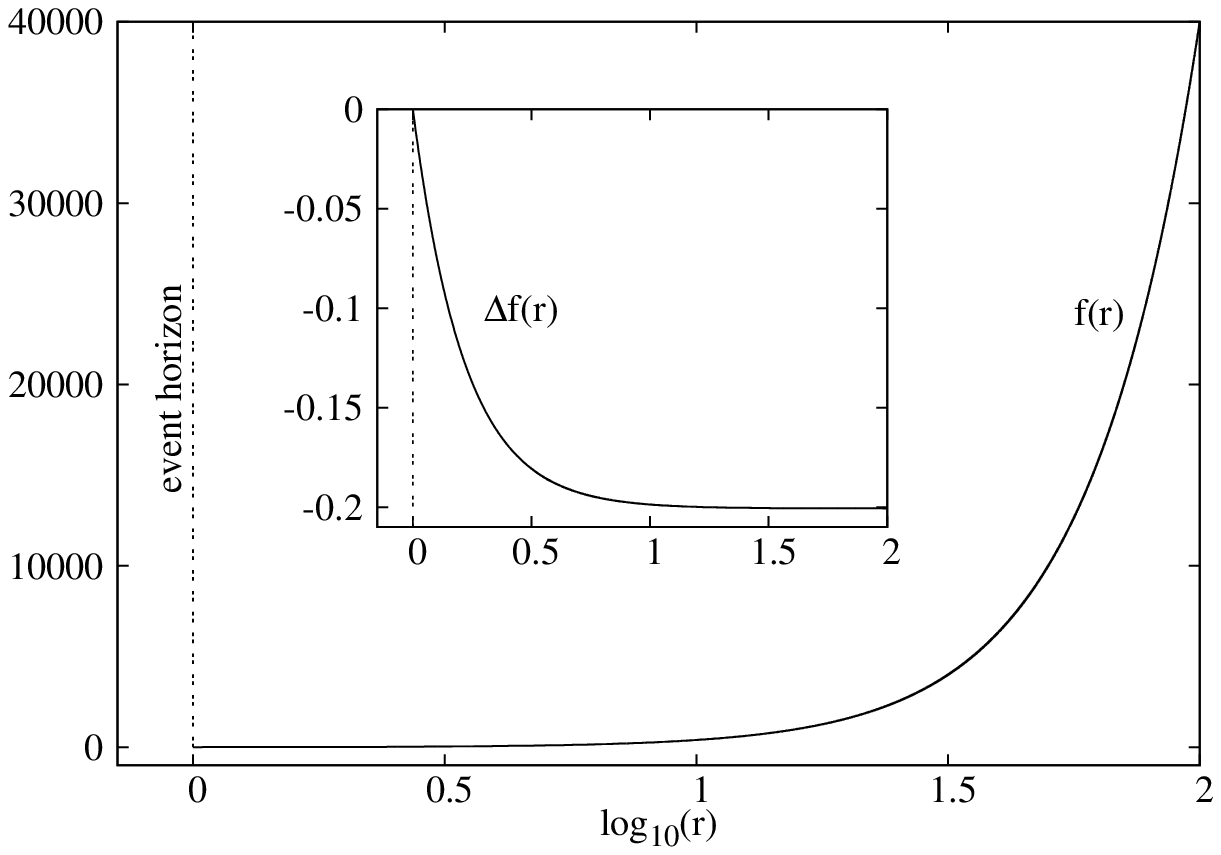}}	
\hss}
 \end{figure}
 %
\vspace*{-0.1cm}
 {\small \hspace*{3.cm}{\it  } }
\begin{figure}[ht]
\hbox to\linewidth{\hss%
	\resizebox{8cm}{6cm}{\includegraphics{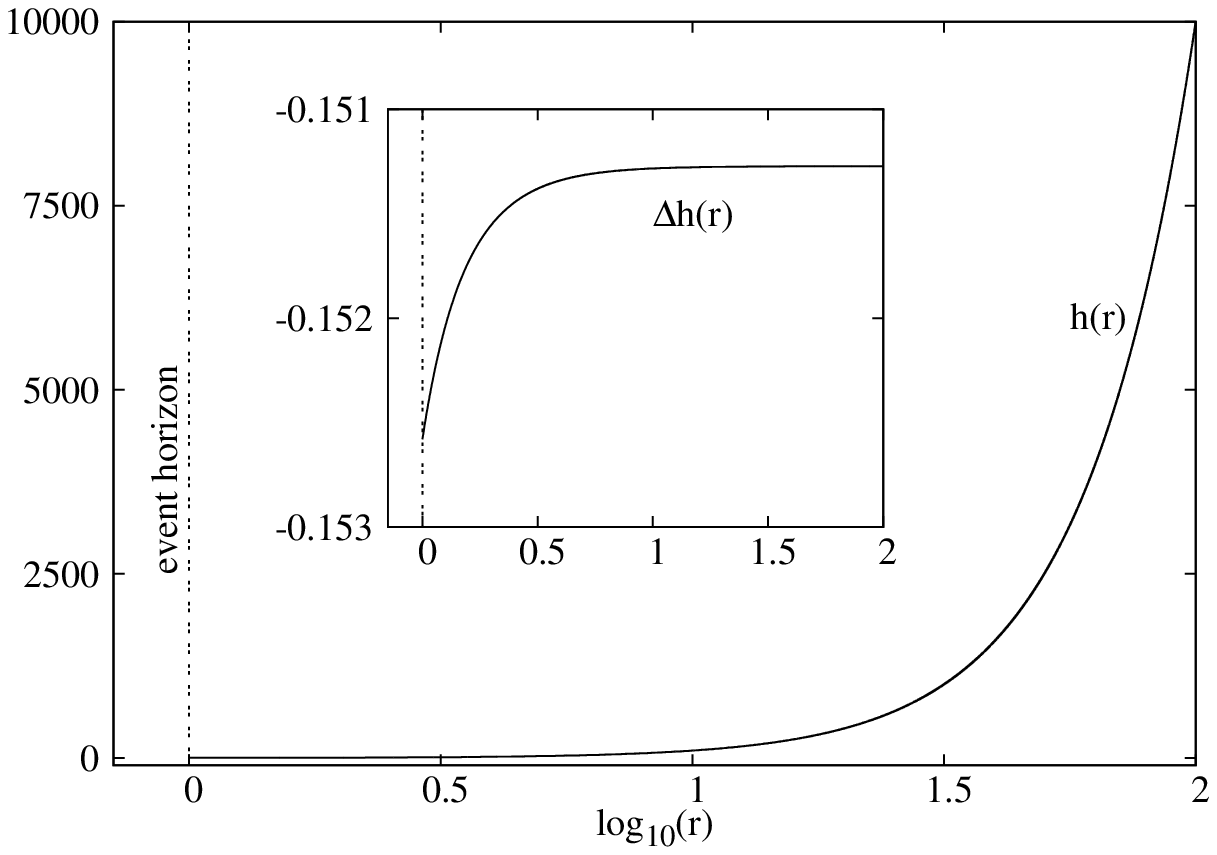}}
\hspace{10mm}%
        \resizebox{8cm}{6cm}{\includegraphics{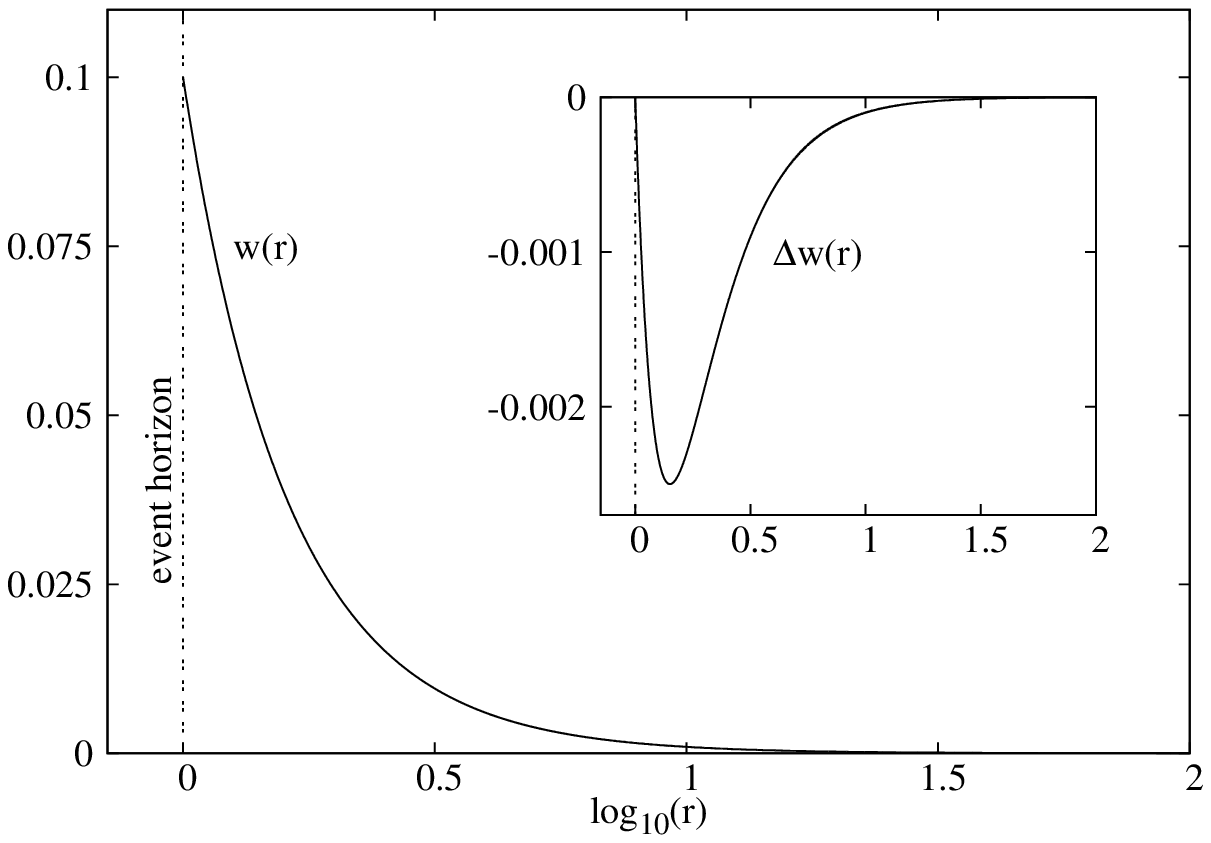}}	
\hss}
 \caption{\small
The profiles of a typical first branch solution with $r_H=1$, $\Omega_H=0.1$ and $\ell=0.707$ 
are shown as a function of the radial coordinate.
The insets show the difference between this solutions and the corrresponding second branch configuration
($e.g.$ $\Delta b(r)=b^{1st}(r)-b^{2nd}(r)$).
  }
\end{figure}
\\
 One can see that 
   the term $2r^2/\ell^2$ 
starts dominating  the profiles of  $b,f$ and $h$ very rapidly.
Also, typically,
the metric functions interpolate monotonically between the corresponding values 
at $r=r_H$ and the asymptotic values at infinity, without presenting 
any local extrema.  

We mention also that similar to other rotating solutions,
these black present also an ergoregion 
inside of which the observers cannot remain stationary,
and will move in the direction of rotation.
 The ergoregion is the region bounded by the event horizon,
 and the stationary limit surface, or the ergosurface, $r = r_e$. 
 The Killing vector $\partial/\partial_t$
becomes null on the ergosurface , i.e. $g_{tt}(r_e) = -b(r_e)+h(r_e)w^2(r_e) = 0$. 
The ergosurface does not intersect
the horizon.

A feature of these solutions which can already be anticipated
based on the perturbative results  is their nonuniqueness in terms of $(r_H,\Omega_H)$
(this contrasts with the case of MP-AdS 

\vspace*{-0.1cm}
 {\small \hspace*{3.cm}{\it  } }
\begin{figure}[ht]
\hbox to\linewidth{\hss%
	\resizebox{8cm}{6cm}{\includegraphics{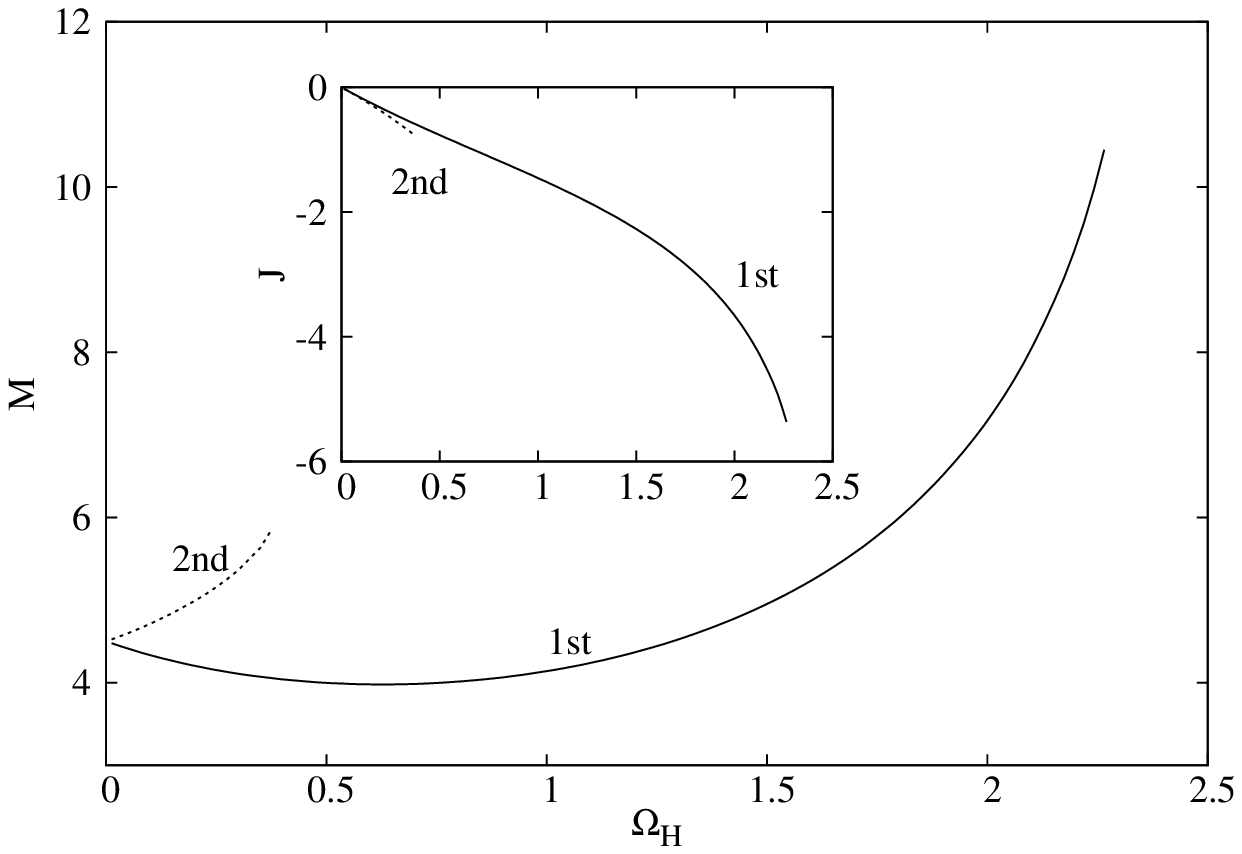}}
\hspace{10mm}%
        \resizebox{8cm}{6cm}{\includegraphics{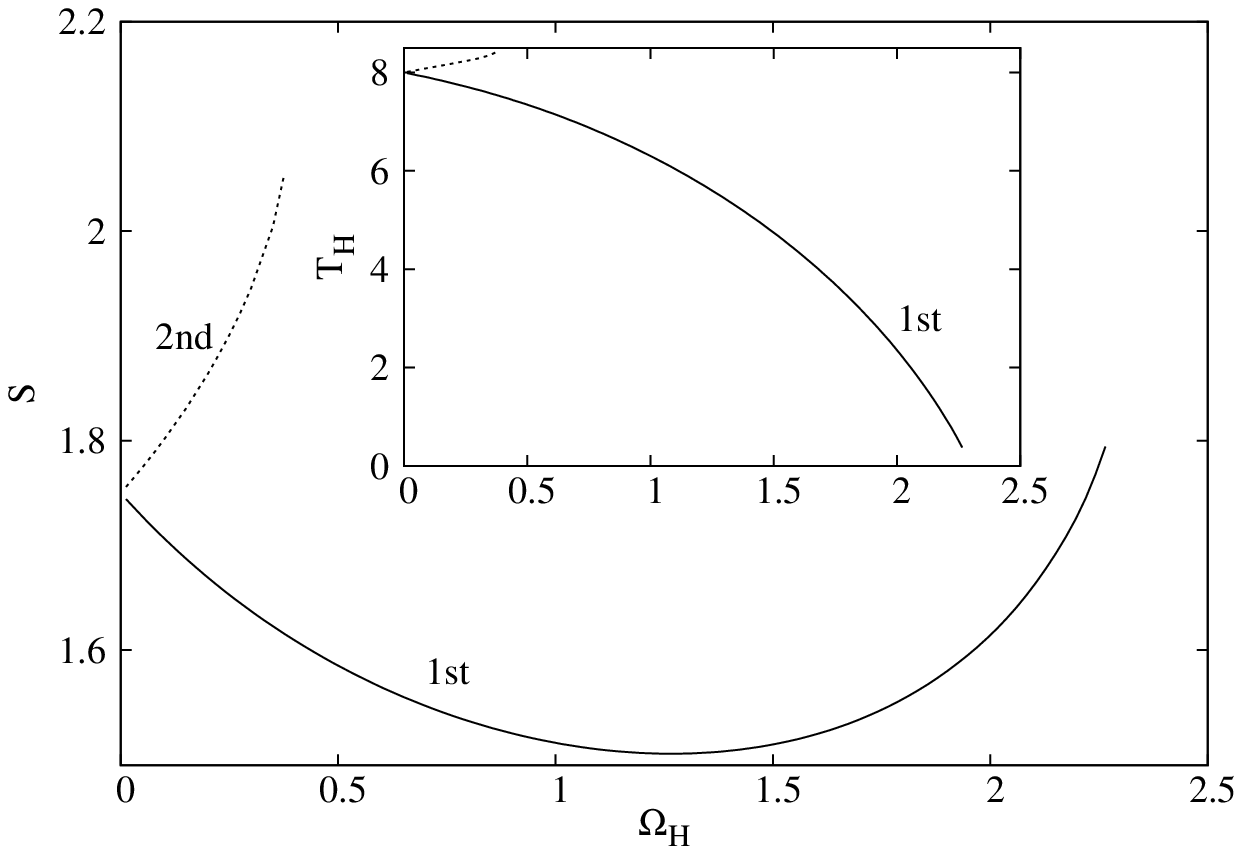}}	
\hss}
 \caption{\small
The mass ${\cal M}$, angular momentum $J$,
 entropy $S$ and Hawking temperature $T_H$ are shown $vs.$
 the angular velocity of the horizon for 1st and 2nd branch solutions
 with $r_H=1$ and $\ell=0.707$ 
 (here and in Figure 3 we set $G=1$).
  }
\end{figure}

 {\small \hspace*{3.cm}{\it  } }
\begin{figure}[ht]
\hbox to\linewidth{\hss%
	\resizebox{8cm}{6cm}{\includegraphics{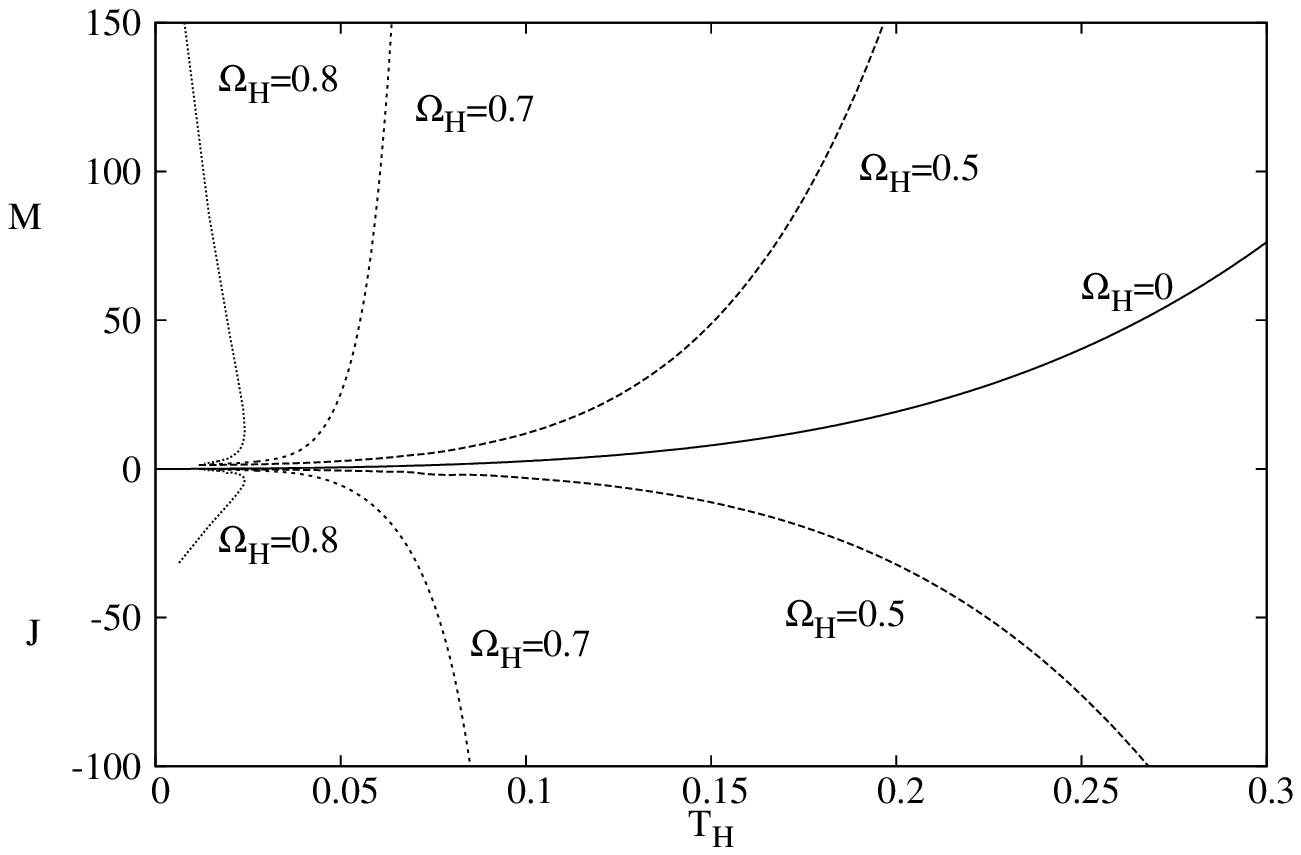}}
\hspace{10mm}%
        \resizebox{8cm}{6cm}{\includegraphics{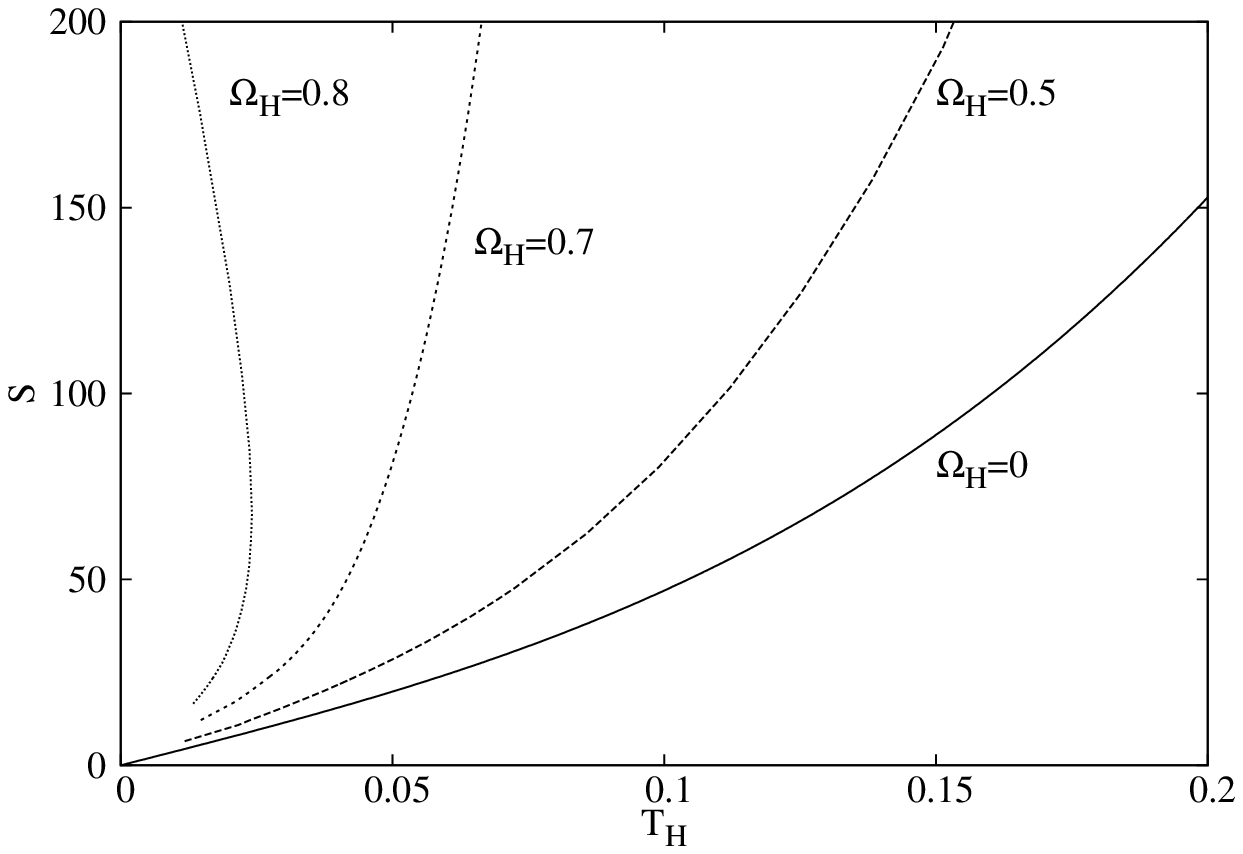}}	
\hss}
 \caption{\small
The mass ${\cal M}$, angular momentum $J$ and
 entropy $S$ are shown as a function of
 the Hawking temperature $T_H$  for 1st branch solutions with
 $\ell=2$ 
and several fixed values of 
 the angular velocity of the horizon.
   }
\end{figure}
 \\
 black holes).
That is, given a static black hole solution with some $r_H>0$,
we have found numerical evidence for the occurance
of two different branches of spinnning solutions 
with the same value of the event horizon velocity $\Omega_H$.
The profiles in Figure 1 correspond to a 1st branch solution.
The insets there show the difference between and first and second branch solutions.
  The two branches can be distinguished, namely, by the value $k(r_H)$ (where we define $k(r) \equiv h(r)/r^2$).
On one branch, the solutions have $k(r_H) < 1, k'(r_H) > 0$  (we will 
 refer to it as the first branch); the second branch
 has $k(r_H) > 1, k'(r_H) < 0$. In relation with the perturbative expansion (\ref{sol-slow})), the first (resp. second) branch
 corresponds to  $P=0$ (resp. $P=1$).

The solutions of the first branch  
obey a standard pattern: 
for a fixed value of the event horizon radius $r_H$,
their temperature decreases when $\Omega_H$ increases. Accordingly, a configuration
with an extremal horizon at $r=r_H$ is approached for a maximal
event horizon velocity, say $\Omega_H = \Omega_{H,ex}$. This value, of course,
depends on $r_H$, $\ell$. Setting $r_H =1$, we  find for example
$\Omega_{H,ex}\approx 2.35$ for $\ell^2 = 1/2$ (case of Figure 2)  
and  $\Omega_{H,ex}\approx 0.935$ for $\ell^2 = 2$.
 
The solutions of the second branch 
exhibit a very different picture.
For example, rather unusual, their temperature increase with $\Omega_H$.
Also, we have found that they have a relatively small extension in $\Omega_H$.
  In fact, increasing of the event horizon velocity result in a very rapid increases
of the pameter $|k'(r_H)|$, suggesting that  
a critical configuration is approached for some $\Omega_H^{max}$.
Although the numerical accuracy is lost as $\Omega_H \to \Omega_H^{max}$,
our results indicate that the horizon becomes singular for the
critical configuration.
Some of these features are illustrated in Figure 2, where we show a number physical
quantities for 1st and 2nd branch solutions as a function of $\Omega_H$
(there we take $r_H=1$ and $\ell=0.707$; however, similar results have been found for 
other choices of these paramters).

Returning on 1st branch solutions (which are likely to be physically more interesting),
we show in Figure 3 the mass, angular momentum and entropy of solutions as  functions of temperature
for several fixed values of $\Omega_H$ (thus we consider configurations 
in a grand canonical ensemble).
The control parameter there is the event horizon radius, which varies between a small value
close to zero (although this region is difficult to explore numerically)
up to a maximal value  which depends on $\Omega_H$.
The results in Figure 3 suggest the following picture.
Similar to the static case, 
the spinning solutions appear to possess a $r_H=0$ limit,
which has a vanishing 
temperature\footnote{This limiting configuration could not be studied numerically. 
However, based on the perturbative results,
we expect it to be singular.}.
When increasing the event horizon radius, a branch of spinning black holes -regular on and outside the horizon- 
occurs, the temperature, mass and the absolute value of $J$
increase along the branch.
However, different from the static case, for any $\Omega_H\neq 0$,
there is a maximal value for the temperature which is approached for some critical value of $r_H$.
At that point, a secondary branch of solutions emerges\footnote{For the data plotted in Figure 3,
this can be seen for $\Omega_H=0.8$.
However, the  backbending occurs for other values of $\Omega_H$ there as well, although for  
  larger values of $T_H$, ${\cal M}$ and $|J|$.},
which extends backwards in
$T_H$ (note that the event horizon radius, the black hole entropy and 
the global charges still increase along this branch).
The end point 
of this branch is an extremal configuration with vanishing 
temperature, finite (and novanishing) horizon size and finite global charges.  
 
These limiting extremal configurations, however,  satisfy a different 
set of boundary conditions at $r=r_H$ than the nonextremal ones
(for example $f|_{r\to r_H}={\cal O}(r-r_H)^2$, $b|_{r\to r_H}={\cal O}(r-r_H)^2$).
(Note that the extremal solutions keep the asymptotic form (\ref{BH-rot-inf}), 
with a single essential parameter in the expansion there.)
Finding such solutions explicitly or numerically 
is beyond the scope of this paper.

\subsection{Extremal solutions: near horizon geometry 
and the entropy function}

However, as usual, some information about the properties of the  
extremal black holes can be obtained by studying the near horizon solution together with the 
corresponding  entropy function.
 
Therefore we consider the following metric form,
describing  
a rotating squashed $AdS_2\times S^3$ spacetime
\begin{eqnarray}
\label{at1}
ds^2=v_1(\frac{dr^2}{r^2}-r^2 dt^2)+\frac{v_2}{4}(\sigma_1^2+\sigma_2^2)
+\frac{v_2v_3}{4}(\sigma_3+2 k r dt)^2
\end{eqnarray}
($i.e.$ for $b(r)=v_1 r^2$, $f(r)=r^2/v_1$, $g(r)=v_2$,
$h(r)=v_2 v_3$, $w(r)=k r$ within the parametrization  (\ref{metric2})),
such that the horizon is located\footnote{This position of the horizon can always be obtained by taking
$r\to r-r_H$.} at $r=0$.
This geometry describes a fibration of $AdS_2$ over the homogeneously squashed $S^3$ with symmetry
group $SO(2, 1)\times SU(2)\times U(1)$ \cite{Kunduri:2007qy}.
Also, (\ref{at1}) would correspond to the neighbourhood of the event horizon of an
 extremal limit of the spinning black holes discussed above.

Given this ansatz, the equations  (\ref{eqs}) of the model reduce to
a set of algebraic relations for the 
 parameters $v_i,k$. 
In what follows we choose to determine these relations by using the  
formalism proposed in \cite{Astefanesei:2006dd}, thus by
 extremizing  an entropy function.
This allows us also to compute the entropy of the extremal black holes 
and to show that these CS solutions exhibit attractor behaviour.
 
Therefore, as usual in the literature, 
let us denote by $f(k,\vec v)$ the lagrangian density $\sqrt{-g} {\cal L}$ 
(as read from (\ref{EGB0}) together with (\ref{cond}), (\ref{alpha}))
evaluated for the near horizon geometry (\ref{at1}),
and integrated over the angular coordinates,
\begin{eqnarray}
\label{at2}
f(k,\vec v)&=&\int d\bar \theta d \phi d\psi \sqrt{-g} {\cal L}
=\frac{1}{16 \pi G} \int d\bar \theta d \phi d\psi \sqrt{-g}  (R+\frac{12}{\ell^2} +\frac{\ell^2}{8}L_{GB})  .
\end{eqnarray}
The field equations (\ref{eqs}) for the near horizon geometry (\ref{at1}) 
 now correspond to
$\frac{\partial f}{\partial k}=J,~~\frac{\partial f}{\partial v_i}=0,$
with $J$ the angular momenta of the solutions.

Then,
following  \cite{Astefanesei:2006dd}, we define the entropy function by taking the Legendre transform of the above integral
with respect to the parameter $k$,
\begin{eqnarray}
\label{at4}
{\cal E}(J,k,\vec v)=2 \pi (J k-f(k,\vec v)).
\end{eqnarray}
The entropy and the near horizon geometry of the spinning black holes
are obtained by extremizing this entropy function.
This leads to the algebraic equations of motion 
  \begin{eqnarray}
 \label{at32}
 \frac{\partial {\cal E}}{\partial k}=0,~~\frac{\partial {\cal E}}{\partial v_i}=0, 
  \end{eqnarray} 
 the entropy associated with the  black hole being
$S_{extremal}={\cal E}(J,\vec v)$
evaluated at the extremum (\ref{at32}). 
 
For the metric ansatz (\ref{at1}), a straightforward calculation gives
\begin{eqnarray}
\label{at21}
{\cal E}(J,\vec v)=
2\pi \bigg [
Jk -\frac{\pi \sqrt{v_2 v_3}}{16 G v_1}
\bigg(
\frac{24v_1^2v_2}{\ell^2}
+
k^2 v_2^2 v_3-4 v_1^2 (v_3-4)-4 v_1 v_2
\\
\nonumber
-\frac{\ell^2}{2} (k^2 v_2 v_3(3v_3-4)-4v_1(v_3-4) )
\bigg)
\bigg ].~~{~~}
\end{eqnarray}
Then
the explicit form of the  equations (\ref{at32}) is
\begin{eqnarray}
\label{at5v}
&&
 -16 v_1^2+4 v_1^2 v_3+k^2 v_2^2 v_3+
\frac{\ell^2}{2} k^2 v_2v_3(4-3v_3)-\frac{24}{\ell^2}v_1^2v_2=0,
\\
\nonumber
&&
-16 v_1^2+12 v_1 v_2+4 v_1^2v_3-5k^2 v_2^2 v_3+
\frac{\ell^2}{2}(-4v_1(v_3-4)+3k^2v_2 v_3(3v_3-4))-\frac{72}{\ell^2} v_1^2v_2=0,
\\
\nonumber
&&
-16 v_1^2+4 v_1 v_2+12 v_1^2v_3-3k^2 v_2^2 v_3+
\frac{\ell^2}{2}(-4v_1(3v_3-4)+3k^2v_2 v_3(5v_3-4)) -\frac{24}{\ell^2}v_1^2v_2=0,
\end{eqnarray}
together with
\begin{eqnarray}
\label{at5}
J=\frac{\pi k (v_2 v_3)^{3/2}}{8 G v_1}
( v_2+\frac{\ell^2}{2}(4-3v_3) ).
 \end{eqnarray} 
 
 A solution of the above equations 
takes a relatively simple form when expressed in terms of the 
'relative squashing' parameter $v_3$.
One finds
 \begin{eqnarray}
\label{at7}
&&v_1=\frac{\ell^2}{8}\frac{(6v_2+\ell^2(4-v_3))(2v_2+\ell^2(4-3v_3)}
{24 v_2^2+\ell^4(4-v_3)(4-3v_3)-6\ell^2v_2(5v_3-8)},
\\
\label{at71}
&&J=\frac{\pi v_2 v_3}{4G}\sqrt{(4-v_3+\frac{6v_2}{\ell^2})(v_2+ \ell^2   (2-\frac{3v_3}{2}))},~~{~~}
\end{eqnarray}
and 
 \begin{eqnarray}
\label{at8}
k=\frac{16 G J v_1}{\pi (v_2 v_3)^{3/2}(4\ell^2+2v_2-3\ell^2v_3 )}.
\end{eqnarray}
The radius $v_2$ of the round $S^2$ sphere is also a function of $v_3$, with\footnote{One can notice that the relations (\ref{at7})-(\ref{at9})
are invariant under the scaling
$
  v_1 \to \lambda v_1,~
v_2 \to \lambda v_2, ~
{\cal E} \to \lambda^{3/2} ~
{\cal E} , ~
J \to \lambda^{3/2} J,~
k \to k,~{\rm and}~\ell \to \lambda^{1/2} \ell,
$
which shows that the solutions exist for any $\ell$.} 
  \begin{eqnarray}
\label{at9}
 v_2=\frac{\ell^2}{2}
 \left(
 3v_3-2+\sqrt{4+8(v_3-1)v_3}
 \right ).
\end{eqnarray}

Finally, inserting these expressions into Eq.~(\ref{at21})
 we obtain for the entropy function of the
extremal black hole:
 \begin{eqnarray}
\label{at10}
{\cal E}=S_{extremal}=  \frac{\pi^2}{2G}\sqrt{v_2 v_3}\left(v_2-\frac{\ell^2}{2}(v_3-4)  \right),
\end{eqnarray}
 (with $v_2(v_3)$ as implied by (\ref{at9})).
We have verified that the above result  
agrees with Wald's form (\ref{S-Noether}) 
evaluated for the near horizon geometry (\ref{at1}).
We also note that, in principle, ${\cal E}$ can be expressed in terms of the conserved
charge $J$ by inverting the relation (\ref{at71}) together with (\ref{at9}).
 
We interpret these results as an indication for the existence of corresponding bulk extremal solutions,
in which case, for the parametrization in this work, $v_2=r_H^2$ and $v_3=h_H/r_H^2$.

\section{Conclusions}
The main purpose of this paper was to
discuss the basic properties of several types of 
black object solutions in $d=5$ Chern-Simons-AdS theory of gravity.
Apart from the known Schwarschild black holes, we  
have considered also
squashed black holes and 
uniform black strings.
Remarkably, different from the case of pure Einstein gravity,
the expression of  the squashed black holes and  
uniform black strings could be found in closed form.
Our results show that the properties of the known 
Schwarschild black holes in CS theory are somehow generic.

Apart from that, 
we have given arguments for the existence of 
rotating black holes in $d=5$ Chern-Simons theory of gravity.
These configurations posses
a regular horizon of spherical topology and 
have two equal-magnitude angular momenta, representing generalizations
of a particular class of Myers-Perry black holes.
  The structure of the asymptotic series suggests 
   the existence of an exact solution also in that case, but so far we could not find it.
Therefore we have studied the properties of the  spinning solutions
using both analytical
and numerical methods.
Exact solutions were constructed
in the slowly rotation limit,
 by considering them
 as a perturbations around the Schwarzschild black hole. 
The nonperturbative solutions were found by solving numerically the field equations.
We hope that these results will prove useful
in constructing  the solutions in a closed form.

We also proposed to adapt the boundary counterterm formalism of 
\cite{Balasubramanian:1999re} to the CS-AdS theory, computing in this way
the global charges  of all solutions in this work. 
This formalism can also be generalized to  the situation when  matter fields are added to the bulk action  (\ref{EGB0}).
For example, we have verified verify that the counterterms 
proposed in Section 2 regularize  the action and mass of the  Reissner-Nordstrom
generalizations of the black holes (\ref{SGB}).  
  
As avenues for further research, 
it would be interesting to consider other classes of solutions apart from those in this work.
Here let us mention that we have also found 
 a class of exact solutions that
 can be viewed as the $d = 4$ Taub-NUT-AdS solutions uplifted to five dimensions,
in the presence of a negative cosmological constant 
(the corresponding Einstein gravity configurations are discussed in \cite{Brihaye:2009dm}).
 Moreover, the present work could also be extended by supplementing 
 the model (\ref{EGB0}) with matter fields,
 for instance gauged scalar fields. This would lead to  charged and/or hairy generalisations of the
 black holes solutions that we have presented. 
 Such solutions were obtained $e.g.$ in
\cite{Brihaye:2012ww} for the generic values of the Gauss-Bonnet coupling constant.
Finally, let us mention that we expect the results in this work to admit direct generalizations to the
higher dimensional case $d=2n+1\geq 7$.

\vspace*{0.7cm}

\noindent
{\bf\large Acknowledgements}\\
The authors would like to thank R. Olea for collaboration in the initial stages of this
project and for further valuable discussions.
Also, E.R. would like to thank the physics department of the Aveiro University for 
the hospitality while this work was in progress.
E.R. gratefully acknowledges support by the DFG,
in particular, also within the DFG Research
Training Group 1620 ``Models of Gravity''.

\newpage

\begin{appendix}
\setcounter{equation}{0}

\section{Rotating solutions:  the system of differential equations}

Given the metric ansatz (\ref{metric}), one 
can show that the tensor $E_{\mu }^\nu$ as defined by
 (\ref{eqs}) has only five  linearly independent components: 
$E_r^r$, 
$E_\theta^\theta$, 
$E_{\varphi_1}^{\varphi_2}$, 
$E_{\varphi_1}^{t}$
and
$E_{t}^{t}$.
All other equations are identically zero or
are linear combinations of 
these  components. One finds  $e.g.$
\begin{eqnarray}
\label{rel1}
\cos^2\theta E_{\varphi_1}^t=\sin^2\theta E_{\varphi_2}^t,~~
E_{\varphi_1}^{\varphi_1}-E_{\varphi_2}^{\varphi_2}+\frac{\cos2\theta}{\sin^2\theta}E_{\varphi_1}^{\varphi_2}=0,
E_\theta^\theta-E_{\varphi_1}^{\varphi_1}+E_{\varphi_2}^{\varphi_2}=0.
\end{eqnarray}
The  explicit expression of the essential components of $E_{\mu }^\nu$  is
\begin{eqnarray}
\label{eq1}
E_r^r 
&=&
-\left(1+\frac{2\ell^2}{3r^2}-\frac{\ell^2h}{6 r^4} -\frac{\ell^2f}{6 r^3} \right)\frac{6}{\ell^2}
+\left(1+\frac{\ell^2h}{2 r^4} \right)\frac{fb'}{rb}
+\frac{fh'}{rh}
\\
\nonumber
&&
+\left(1+\frac{2\ell^2}{3r^2}-\frac{3\ell^2h}{2 r^4} -\frac{3\ell^2f}{2 r^3} \right)
\left(\frac{b'h'}{h}+hw'^2 \right)\frac{f}{4b},
\end{eqnarray}
\begin{eqnarray}
\label{eq2}
E_\theta^\theta
&=&
-\frac{6}{\ell^2}
-\frac{h}{r^4}
+\left(
\frac{b'}{b}+\frac{f'}{f}
\right) \frac{f}{2r}
+\left(
1+\frac{rb'}{b}
\right)\frac{fh'}{2rh}
-\left(
1-\frac{\ell^2fh'}{4rh}+\frac{\ell^2h}{2r^4}
\right)\frac{fb'^2}{4b^2}
\\
\nonumber
&&
+\left(
1-\frac{\ell^2fh'}{2rh}+\frac{\ell^2h}{2r^4}
\right)\frac{b'f'}{4b}
+\left(
1+\frac{3\ell^2h}{2r^4}-(\frac{b'}{b}-\frac{3f'}{f}-\frac{3h'}{h})\frac{\ell^2f}{4r}
\right )\frac{fh}{4b}w'^2
\\
\nonumber
&&
+\left(
1-\frac{\ell^2fh'}{4rh}+\frac{\ell^2h}{2r^4}
\right)\frac{fb''}{2b}
+\left(
1-\frac{\ell^2 f b'}{4rb}
\right)
\left(h''-\frac{h'^2}{2h}+\frac{f'h'}{2f} \right)
\frac{f}{2h}
-\frac{\ell^2f^2h}{4rb}w'w'',
\end{eqnarray}
\begin{eqnarray}
\nonumber
\frac{1}{\sin^2\theta}E_{\varphi_1}^{\varphi_2} 
&=& 
\left(
\frac{f'}{f}+\frac{b'}{b}
\right)\frac{f}{2r}
-\left(
1-\frac{f}{4}-\frac{h'}{r^2}
\right)
\left(
1+\frac{\ell^2f b'^2}{8b^2}
\right)\frac{4}{r^2}
-\left(
1+\frac{rf'}{2f}-\frac{rh'}{2h}
\right)\frac{fh'}{2rh}
\\
\label{eq3}
&&
+\left(
1-\frac{3f}{4}-\frac{h}{r^2}
\right)\frac{\ell^2b'f'}{2r^2b}
-\left(
1+
(\frac{h'}{h}+\frac{b'}{b}-\frac{3f'}{f})\frac{\ell^2f}{4r}
\right)
\frac{fb'h'}{4bh}
\\
\nonumber
&&
-\left(
1+\frac{3\ell^2h}{2r^4}-\frac{\ell^2}{4r}(f'+\frac{3h'}{r^2})
\right)
\frac{fh}{rb}ww'
-\bigg(
1+\frac{3\ell^2}{r^2}-\frac{3\ell^2 f}{4r^2}-\frac{9\ell^2h}{2r^2} 
\\
\nonumber
&&
+\frac{3\ell^2f}{8r}(\frac{b'}{3b}-\frac{f'}{f}-\frac{h'}{h})
\bigg)\frac{fh}{2b}w'^2
+\left(
1-\frac{f}{4}-\frac{h}{r^2}+\frac{rfh'}{8h}
\right)\frac{\ell^2f b''}{r^2b}
-\left(
1-\frac{\ell^2fb'}{4rb}
\right)\frac{fh''}{2h}
\\
\nonumber
&&
-\left(
1+\frac{2\ell^2}{r^2}-\frac{\ell^2f}{2r^2}-\frac{3\ell^2 h}{2r^4}
\right)
\left(
(-\frac{b'}{b}+\frac{f'}{f}+\frac{3h'}{h})\frac{w'}{2}+w''
\right)\frac{wfb}{2b}
+\frac{\ell^2f^2 h}{4rb}w'w'',
\end{eqnarray}
\begin{eqnarray}
\label{eq4}
\frac{1}{\sin^2\theta}E_{\varphi_1}^{t} 
&=& 
\bigg[
-\left(1+\frac{3\ell^2h}{2r^4} \right)\frac{fh}{rb}
+\left(
1+\frac{2\ell^2}{r^2}-\frac{\ell^2 f}{2r^4}-\frac{3\ell^2h}{2r^4}
\right)\frac{fhb'}{4b^2}
\\
\nonumber
&&
\left(
1+\frac{2\ell^2}{r^2}-\frac{3\ell^2 f}{2r^4}-\frac{3\ell^2h}{2r^4}
\right)\frac{ hf'}{4b }
-\left(
1+\frac{2\ell^2}{r^2}-\frac{3\ell^2f}{2r^2}-\frac{5\ell^2 h}{2r^4}
\right)\frac{3fh'}{4b }
\bigg ]w'
\\
\nonumber
&&
-\left(
1+\frac{2\ell^2}{r^2}-\frac{ \ell^2f}{2r^2}-\frac{3\ell^2 h}{2r^4}
\right)
\frac{fh}{2b}w'',
\end{eqnarray}
\begin{eqnarray}
\nonumber
 E_{t}^{t} 
&=& 
-\frac{4}{\ell^2}
-
\left(
1+\frac{2\ell^2}{r^2}-\frac{ \ell^2f}{2r^2}-\frac{ \ell^2 h}{2r^4}
\right)\frac{2}{\ell^2}
+\frac{f'}{r}
-\left(
1-r(\frac{f'}{6f}+\frac{h'}{h})
\right)\frac{3\ell^2fh}{r^4}
+\frac{fh'}{rh}
\\
\label{eq5}
&&
\left(
1+\frac{3\ell^2 h}{2r^4}
\right)\frac{fh}{rb}ww'
+\left(
1+\frac{2\ell^2}{r^2}-\frac{3 \ell^2f}{2r^2}-\frac{ 3\ell^2 h}{2r^4}
\right )
\left(\frac{h'}{h}+\frac{hww'}{b} \right)\frac{f'}{4}
\\
\nonumber
&&
-\left(
1+\frac{2\ell^2}{r^2}-\frac{ \ell^2f}{2r^2}+\frac{ 3\ell^2 h}{2r^4}
\right )
\frac{fh'^2}{4 h^2}
+\left(
1+\frac{2\ell^2}{r^2}-\frac{ \ell^2f}{2r^2}-\frac{ 5\ell^2 h}{2r^4}
\right )
\frac{3fh' }{4b}ww'
\\
\nonumber
&&
+\left(
1+\frac{2\ell^2}{r^2}-\frac{ \ell^2f}{2r^2}-\frac{ 3\ell^2 h}{2r^4}
\right )
\left(
\frac{h''}{h}-\frac{wb'w'}{2b^2}+\frac{w'^2}{2b}+\frac{ww''}{b}
\right)\frac{fh}{2}.
\end{eqnarray}
However, the following relation holds\footnote{The existence of (\ref{id1}) is a consequence of the generalized Bianchi
identities $E_{\mu;\nu}^\nu=0$.}: 
\begin{eqnarray}
\label{id1}
\frac{d E_r^r}{dr}&=&\left(-\frac{2}{r}-\frac{b'}{2b}-\frac{h'}{2h} \right)E_r^r
+\left(
\frac{2}{r}+\frac{h'}{2h}
\right)E_\theta^\theta
+\frac{h'}{2h}\frac{E_{\varphi_1}^{\varphi_2}}{\sin^2\theta}
\\
\nonumber
&&
\left(
\frac{1}{2}w(\frac{b'}{b}-\frac{h'}{h}-w')
\right)\frac{E_{\varphi_1}^{t}}{\sin^2\theta}
+\frac{b'}{2b}E_t^t~.
\end{eqnarray}
Thus we are left with a set of four essential equations for the metric functions
$f,b,h,w$.
Note also that
the equation $E_{\varphi_1}^t$ implies the existence of the first integral
\begin{eqnarray}
\label{fi-rotation}
r^2h\sqrt{\frac{fh}{b}}
\left(
1+\frac{2\ell^2}{r^2}-\frac{ \ell^2f}{2r^2}-\frac{ 3\ell^2 h}{2r^4}
\right)w'=const.
\end{eqnarray}

In the numerical construction of the solutions, we have choosen to solve a suitable linear combination of the equations
\begin{eqnarray}
\label{set}
E_{\theta}^\theta=0,~~
E_{\varphi_1}^{\varphi_2}=0,~~
 E_{\varphi_1}^{t}=0 ~~
{\rm and}~~
 E_t^{t}=0.
\end{eqnarray}
The remaining equation $E_{r}^r=0$  
becomes a constraint, which is implicitly satisfied for the choosen 
set of boundary conditions.
Here we use the observation that (\ref{set}) implies that (\ref{id1}) can be written as
$\frac{d}{dr}(r^2 \sqrt{bh} E_r^r)=0$, $i.e.$ $r^2 \sqrt{bh} E_r^r=const$.
However, since $b(r_H)=0$, we find that $E_r^r\equiv 0$.
In practice, we have used the $E_r^r$ equation together with the first integral 
(\ref{fi-rotation})
to monitor the accuracy of the numerical results.

For completness, we mention that the same set of equations are found when considering instead 
an effective lagrangeean approach.
Without fixing a metric gauge, a straightforward computation
leads to the following reduced action for the system
 \begin{eqnarray}
\label{Leff} 
\nonumber
A_{\rm eff}=\int dr
 \bigg[
&&
\sqrt{\frac{fh}{b}}
\bigg(
b'g'+\frac{g}{2h}b'h'+\frac{b}{2g}g'^2+\frac{b}{h}g'h'+\frac{1}{2}gh w'^2+\frac{2b}{f}(4-\frac{h}{g}
+\frac{12}{\ell^2}\frac{bg}{f})
\bigg)
\\
\nonumber
&&
+\frac{\ell^2}{8}
\left(
\frac{4h}{g}b'g'+2(4g-3h)(\frac{b'h'}{h}+hw'^2)
-\frac{f}{2h}b'h'g'^2-\frac{1}{2}fhg'^2w'^2
\right)
\bigg ]~.
\end{eqnarray}
Then, when taking the variation
of $A_{\rm eff}$ with respect to $a$, $b$, $f$, $g$ and $w$
and fixing afterwards the metric gauge $g(r)=r^2$
one finds a linear combination of the Eqs. (\ref{eq1})-(\ref{eq5}). 
 

\section{The coefficients of the perturbative rotating black hole solution}
\setcounter{equation}{0}

The coefficients which enters the perturbative solution (\ref{sol-slow})  
have the following expression, up to order four:
\begin{eqnarray}
\nonumber
&&
b_{11}=0,~b_{21}=-\frac{3}{16}+\frac{\beta^2}{8}-\beta^4,~~b_{22}=0,
\\
\nonumber
&&
b_{31}=\frac{23}{8}-\frac{59}{36}\beta^2+\frac{7}{18}\beta^4-3\beta^6,~~
b_{32}=\frac{1}{576}(387-4\beta^2(305-323\beta^2+432\beta^4)),~b_{33}=0,
\\
\nonumber
&&
b_{41}=\frac{1}{18432}
\left(
-751203+2\beta^2(190027+2\beta^2(-56497+55106\beta^2-57600\beta^4+9216\beta^6))
\right),
\\
\nonumber
&&
b_{42}=\frac{1}{18432}
\left(
-235989+2\beta^2(195277+2\beta^2(-90391+80462\beta^2-63360\beta^4+9216\beta^6))
\right),
\\
\nonumber
&&
b_{43}=\frac{1}{18432}
\left(
-54081+228786\beta^2-305132\beta^4+21644\beta^6-142848\beta^8
\right),~~b_{44}=0,
\end{eqnarray}
\begin{eqnarray}
\nonumber
&&
f_{11}=-\frac{1}{2},~~~f_{21}=\frac{1}{16}(21-22\beta^2),~~~f_{22}=\frac{1}{16}(9+2\beta^2),
\\
\nonumber
&&
f_{31}= \frac{1}{576}(-6039+5020\beta^2-3292\beta^4+288\beta^6),~~
f_{32}=\frac{1}{288}(-1683+308 \beta^2+52 \beta^4),
\\
\nonumber
&&
f_{33}= \frac{1}{576}(-2043+20 \beta^2(17+  \beta^2)), 
\\
\nonumber
&&
f_{41}= \frac{1}{6144}(724131-598870\beta^2+391492 \beta^4-191496 \beta^6+35328 \beta^8),
\\
\nonumber
&&
f_{42}=\frac{1}{18432}(1277973-2\beta^2(182029-83630 \beta^2+5020 \beta^4+2304 \beta^6)),
\\
\nonumber
&&
f_{43}=\frac{1}{18432}(898299-285446\beta^2+136420\beta^4-2120\beta^6) ,
\\
\nonumber
&&
f_{44}=\frac{1}{18432}(406575-67502\beta^2-300\beta^4-40\beta^6),
\end{eqnarray}
\begin{eqnarray}
\nonumber
&&
h_{11}=-\frac{3}{2},~h_{21}=\frac{9}{2}-\frac{3\beta^2}{2}+\beta^4,~~h_{22}=\frac{69}{16} +\frac{5\beta^2}{8},
\\
\nonumber
&&
h_{31}= \frac{1}{64}(-2565+836\beta^2-596\beta^4+288\beta^6),~~ 
h_{32}=  \frac{1}{64}(-1863+4\beta^2-284\beta^4),~~
\\
\nonumber
&& 
h_{33}=  \frac{1}{192}(-4311-316\beta^2+100\beta^4),~~ 
\\
\nonumber
&&
h_{41}=\frac{1}{1536}
(
730899-256454\beta^2+199012\beta^4-97736\beta^6+37248\beta^8-3072\beta^{10}
) ,
\\
\nonumber
&&
h_{42}= \frac{1}{1024}
(
325629-21066\beta^2+52476\beta^4-17528\beta^6+1280\beta^8 
) ,
\\
\nonumber
&&
h_{43}=  \frac{1}{1536}
(
352845+694\beta^2+60668\beta^4+2952\beta^6 
) ,
\\
\nonumber
&&
h_{44}=  \frac{1}{6144}
(
874665+21310\beta^2-27316\beta^4+2280\beta^6 
) ,
\end{eqnarray}
\begin{eqnarray}
\nonumber
&&
w_{11}=1,~w_{21}= 0,~w_{22}=-1+3\beta^2,
w_{31}=0,~  
w_{32}= 3-\frac{27}{4}\beta^2+\frac{9}{2}\beta^4-2\beta^6,~~
\\
\nonumber
&&
w_{33}= \frac{1}{16}(107+124\beta^2(-1+2\beta^2)),~~ 
w_{41}=0, ~~
\\
\nonumber
&&
w_{42}=  \frac{1}{96}(-2565+6335\beta^2-2844\beta^4+2068\beta^6-864 \beta^8),
\\
\nonumber
&&
w_{43}= \frac{1}{576}(-23733+24798\beta^2-31276\beta^4+16648\beta^6-6912 \beta^8)  ,
\\
\nonumber
&&
w_{44}=  \frac{1}{1152}(-47835+62600\beta^2-39548\beta^4+21266\beta^6 ).
\end{eqnarray}
 
To the same order, the coefficients which enter the expressions (\ref{pt1}) 
of the global quantities are
\begin{eqnarray}
\label{pt2}
 {\cal M}^{(0)}=\frac{3\pi \ell^2}{8G}\beta^2(1+\beta^2),~~
 S^{(0)}=\frac{\pi^2 \ell^3}{4G}\beta (3+2\beta^2),~~
 T_H^{(0)}=\frac{\beta}{\pi \ell}~,
 \end{eqnarray}
 and
\begin{eqnarray}
\label{pt3}
\nonumber
&&
M_1=-\frac{3}{16},~~M_2=\frac{1}{128}(81-18\beta^2+32 \beta^4),~~
M_3=\frac{1}{1536}(-9351+4\beta^2(620-413 \beta^2+456 \beta^4)),
\\
\nonumber
&&
M_4=\frac{1}{49152} (3674799-1054150 \beta^2+872692\beta^4-475272\beta^6
+290304 \beta^8-24576 \beta^{10}),
\\
\nonumber
&&
j_1=3+2\beta^2,~~j_2=\beta^2(1+2\beta^2),~~j_3=\frac{1}{8}\beta^2(1+2\beta^2)(22\beta^2-21),
\\
\nonumber
&&
j_4=-\frac{1}{288}\beta^2(1+2 \beta^2)(-6039+5020 \beta^2-3292 \beta^4+288 \beta^6),
\\
\nonumber
&&
w_1=1,~~w_2=-1+3\beta^2,~~w_3=\frac{1}{16}(155-4\beta^2(58-49\beta^2+8 \beta^4)),
\\
\nonumber
&&
w_4=\frac{1}{1152}\big(-126081+4\beta^2(47054-34057\beta^2+19844\beta^4-6048 \beta^6)\big),
\\
\nonumber
&&
s_1=-3,~s_2=\frac{39}{4}-4 \beta^2+2\beta^4,~
s_3=\frac{1}{96}(-91117+4\beta^2(722-527\beta^2+288\beta^4))),
\\
\nonumber
&&
s_4=\frac{1}{3072}(3589695-2\beta^2(599607-472522\beta^2+235924\beta^4-106752\beta^6+7680\beta^8)),
\\
 \nonumber
&&
t_1=-1,~t_2=\frac{13}{4}-\frac{5}{2}\beta^2-2\beta^4,~
t_3=\frac{1}{288}(-9117+4172\beta^2-1940\beta^4-2880\beta^6),
\\
\nonumber
&&
t_4=\frac{1}{3072}\left(1196565+2\beta^2(-277901+2\beta^2(58871+60002\beta^2-36864\beta^4+5376 \beta^6)) \right).
\end{eqnarray}
We close this part by remarking that 
similar expressions have been found  also for orders five, six and seven, 
without being possible to identify
a general pattern for the coefficients.

\newpage
\section{An exact solution with non-vanishing rotation}
By using an 'educated guess' approach,\footnote{In deriving this solution,
we have started by imposed the $'ad~hoc'$ condition
$
b(r)=\frac{r^2 f(r)}{h(r)},
$
(which holds for the Myer-Perry-AdS$_5$ black hole).
In the next step, we have set $h(r)=r^2U(r)$
and looked for a suitable combination of (\ref{eq1})-(\ref{eq5})
which could be solved analytically.
The last step is to rescale $b$, $w$
to make them compatible with the asymptotics (\ref{BH-rot-inf}). 
}
we have found the following exact solution
of the CS  equations (\ref{eq1})-(\ref{eq5}):
 \begin{eqnarray}
 \label{ex-sol}
&&
f(r)=1+\frac{2r^2}{\ell^2}+ c_1F(r),~~
b(r)= \big(1+\frac{2r^2}{\ell^2}\frac{1}{1+c_1F(r)} \big)(1+c_1),
\\
\nonumber
&&
h(r)=r^2(1+c_1F(r)),~~
w(r)=\frac{\sqrt{2}c_1 }{\ell\sqrt{1+c_1}}\frac{F(r)-1}{1+c_1 F(r)},
\end{eqnarray}
where
\begin{eqnarray}
\nonumber
F(r)=\sqrt{1+\frac{c_2}{r^2}},
\end{eqnarray}
$c_1$ and $c_2>0$ being constants of integration.
As $r\to \infty$, this solution becomes
\begin{eqnarray}
\nonumber
&&
f(r)=\frac{2r^2}{\ell^2}+1+c_1+O(1/r^2),~~
b(r)=\frac{2r^2}{\ell^2}+(1+c_1)(1-\frac{c_1c_2}{(1+c_1)^2\ell^2})+O(1/r^2),
\\
&&
\label{as-exsol}
h(r)=r^2(1+c_1)+\frac{c_1c_2}{2}, ~~w(r)=\frac{c_1c_2}{\sqrt{2}(1+c_1)^{3/2}}\frac{1}{\ell r^2},
\end{eqnarray}
such that the boundary $S^3$ sphere is squashed in this case.
Then  the formalism described in the Section 2 leads to the following global charges of this solution
\begin{eqnarray}
\label{gc-exsol}
{\cal M}=\frac{\pi}{32 G \sqrt{1+c_1}}(-16 c_1^2c_2+(1+c_1)^2(5c_1-3)\ell^2 ),
~~J=\frac{\pi \ell}{8\sqrt{2} G}(1-c_1)c_1 c_2.
\end{eqnarray}
Unfortunately, one can see that this configuration possess some unphysical properties for any 
non-vanishing $c_1,c_2$.
First, if we take $c_1>0$, then $r=0$ corresponds to a naked singularity\footnote{Note that we have failed to find  rotating generalizations
of the $k=1$ AdS background (\ref{SGB}), (\ref{back}),
both numerically and also when looking for a perturbative solution.}
 ($e.g.$ $R\to -3c_1\sqrt{c_2}/r^3$ in that limit). 
If we choose instead to take $c_1<0$  such that the condition $f(r_0)=b(r_0)=0$ is satisfied (with $r_0>0)$, 
one finds
that $r=r_0$ is a singular point,  with $h(r_0)<0$.

Finally, let us mention the existence of another solution, which is find by taking an appropriate limit in (\ref{ex-sol}):
 \begin{eqnarray}
 \label{ex-sol1}
&&
f(r)=1+\frac{2r^2}{\ell^2}+ \frac{c_1}{r},~~
 b(r)= \big(1+\frac{2r^2}{\ell^2}
 \frac{1}{1+\frac{c_1}{r} }\big)(1+c_1),
 \\
 &&
 h(r)=r^2(1+\frac{c_1}{r}),~~
 w(r)=\frac{\sqrt{2}c_1  \sqrt{1+c_1}} { \ell r  (1+\frac{c_1}{r} )}.
\end{eqnarray}
However, one can easily show that this solution inherits all pathologies of (\ref{ex-sol}).
Moreover, its asympotics are nonstandard in this case, with a  different form than (\ref{as-exsol}).

 \end{appendix}
\begin{small}

 \end{small}


\begin{thebibliography}{99}
\bibitem{Lovelock:1971yv}
  D.~Lovelock,
  J.\ Math.\ Phys.\  {\bf 12} (1971) 498.
\bibitem{Zwiebach:1985uq}
  B.~Zwiebach,
  Phys.\ Lett.\ B {\bf 156} (1985) 315.
\bibitem{Boulware:1985wk}
  D.~G.~Boulware and S.~Deser,
  Phys.\ Rev.\ Lett.\  {\bf 55} (1985) 2656.
 
  
\bibitem{Chamseddine:1989nu}
  A.~H.~Chamseddine,
  Phys.\ Lett.\ B {\bf 233} (1989) 291.
\bibitem{Garraffo:2008hu}
  C.~Garraffo and G.~Giribet,
  Mod.\ Phys.\ Lett.\  A {\bf 23} (2008) 1801
  [arXiv:0805.3575 [gr-qc]].
 \bibitem{Charmousis:2008kc}
  C.~Charmousis,
  Lect.\ Notes Phys.\  {\bf 769} (2009) 299
  [arXiv:0805.0568 [gr-qc]].
\bibitem{Jacobson:1993xs}
  T.~Jacobson and R.~C.~Myers,
  Phys.\ Rev.\ Lett.\  {\bf 70} (1993) 3684
  [arXiv:hep-th/9305016].
\bibitem{Wald:1993nt}
  R.~M.~Wald,
  Phys.\ Rev.\  D {\bf 48} (1993) 3427
  [arXiv:gr-qc/9307038].
\bibitem{Fayyazuddin:1998fb}
  A.~Fayyazuddin and M.~Spalinski,
  Nucl.\ Phys.\  B {\bf 535} (1998) 219
  [arXiv:hep-th/9805096].
\bibitem{Aharony:1998xz}
  O.~Aharony, A.~Fayyazuddin and J.~M.~Maldacena,
  JHEP {\bf 9807} (1998) 013
  [arXiv:hep-th/9806159].
\bibitem{Nojiri:2000gv}
  S.~Nojiri and S.~D.~Odintsov,
  JHEP {\bf 0007} (2000) 049
  [arXiv:hep-th/0006232].
  
\bibitem{Maldacena:1997re}
J.~M.~Maldacena,
Adv.\ Theor.\ Math.\ Phys.\  {\bf 2} (1998) 231
[Int.\ J.\ Theor.\ Phys.\  {\bf 38} (1999) 1113]
[arXiv:hep-th/9711200].
  
\bibitem{Cai:1998vy}
  R.~-G.~Cai and K.~-S.~Soh,
  Phys.\ Rev.\ D {\bf 59} (1999) 044013
  [gr-qc/9808067].
\bibitem{Crisostomo:2000bb}
  J.~Crisostomo, R.~Troncoso and J.~Zanelli,
  Phys.\ Rev.\ D {\bf 62} (2000) 084013
  [hep-th/0003271].
\bibitem{Aiello:2004rz}
  M.~Aiello, R.~Ferraro and G.~Giribet,
  Phys.\ Rev.\ D {\bf 70} (2004) 104014
  [gr-qc/0408078].
\bibitem{Murata:2009jt}
  K.~Murata, T.~Nishioka and N.~Tanahashi,
  arXiv:0901.2574 [hep-th].
\bibitem{Brihaye:2009dm}
  Y.~Brihaye, J.~Kunz and E.~Radu,
  JHEP {\bf 0908} (2009) 025
  [arXiv:0904.1566 [gr-qc]].
\bibitem{Copsey:2006br}
  K.~Copsey and G.~T.~Horowitz,
  JHEP {\bf 0606} (2006) 021
  [arXiv:hep-th/0602003].
\bibitem{Mann:2006yi}
  R.~B.~Mann, E.~Radu and C.~Stelea,
  JHEP {\bf 0609} (2006) 073
  [arXiv:hep-th/0604205].
\bibitem{Witten:1988hc}
  E.~Witten,
  Nucl.\ Phys.\ B {\bf 311} (1988) 46.
\bibitem{Chamseddine:1990gk}
  A.~H.~Chamseddine,
  Nucl.\ Phys.\ B {\bf 346} (1990) 213.
\bibitem{Banados:1996hi}
  M.~Banados, R.~Troncoso and J.~Zanelli,
  Phys.\ Rev.\ D {\bf 54} (1996) 2605
  [gr-qc/9601003].
\bibitem{Troncoso:1997va}
  R.~Troncoso and J.~Zanelli,
  Phys.\ Rev.\ D {\bf 58} (1998) 101703
  [hep-th/9710180].
\bibitem{Zanelli:2005sa}
  J.~Zanelli,
  hep-th/0502193.
\bibitem{Okuyama:2005fg}
  N.~Okuyama and J.~i.~Koga,
  Phys.\ Rev.\  D {\bf 71} (2005) 084009
  [arXiv:hep-th/0501044].
\bibitem{Feff-Graham}
  C.~Fefferman and C.~R.~Graham,
  ``Conformal invariants,''
  in {\it \'Elie Cartan et les Math\'ematiques d'Aujourd'hui}, Ast\'erisque,
  (1985), 95-116.  
  
\bibitem{GibbonsHawking1}
G.~W.~Gibbons and S.~W.~Hawking,
  Phys.\ Rev.\  D {\bf 15} (1977) 2752.  
  
\bibitem{Myers:1987yn}
  R.~C.~Myers,
  Phys.\ Rev.\  D {\bf 36} (1987) 392.
\bibitem{Balasubramanian:1999re} 
V.~Balasubramanian and P.~Kraus, 
Commun.\ Math.\ Phys.\ \textbf{208}
(1999) 413 [arXiv:hep-th/9902121]. 

\bibitem{Banados:2004zt}
  M.~Banados, A.~Schwimmer and S.~Theisen,
  JHEP {\bf 0405} (2004) 039
  [hep-th/0404245].
\bibitem{Banados:2005rz}
  M.~Banados, R.~Olea and S.~Theisen,
  JHEP {\bf 0510} (2005) 067
  [hep-th/0509179].

\bibitem{Brihaye:2008kh}
  Y.~Brihaye and E.~Radu,
  Phys.\ Lett.\  B {\bf 661} (2008) 167
  [arXiv:0801.1021 [hep-th]].
\bibitem{Brihaye:2008xu}
  Y.~Brihaye and E.~Radu,
  JHEP {\bf 0809} (2008) 006
  [arXiv:0806.1396 [gr-qc]].
   
\bibitem{Hawking:ig} 
S.~W.~Hawking in \textit{General Relativity. An
Einstein Centenary Survey}, edited by S.~W.~Hawking and W.~Israel,
(Cambridge, Cambridge University Press, 1979).   
\bibitem{Mann:2003-Found} 
R.~B.~Mann,
Found.\ Phys.\ \textbf{33} (2003) 65 [arXiv:gr-qc/0211047].
  \bibitem{Myers:1999ps}
  R.~C.~Myers,
  Phys.\ Rev.\ D {\bf 60}, 046002 (1999)
  [arXiv:hep-th/9903203].
  
\bibitem{Olea:2006vd}
  R.~Olea,
  JHEP {\bf 0704} (2007) 073
  [arXiv:hep-th/0610230].
\bibitem{Mora:2004kb}
  P.~Mora, R.~Olea, R.~Troncoso and J.~Zanelli,
  JHEP {\bf 0406} (2004) 036
  [hep-th/0405267].
\bibitem{Miskovic:2007mg}
  O.~Miskovic and R.~Olea,
  JHEP {\bf 0710} (2007) 028
  [arXiv:0706.4460 [hep-th]].
\bibitem{Kofinas:2007ns}
  G.~Kofinas and R.~Olea,
  JHEP {\bf 0711} (2007) 069
  [arXiv:0708.0782 [hep-th]].
  
   
  
\bibitem{Banados:1992wn}
  M.~Banados, C.~Teitelboim and J.~Zanelli,
  Phys.\ Rev.\ Lett.\  {\bf 69} (1992) 1849
  [hep-th/9204099].
 
 
\bibitem{Dotti:2007az}
  G.~Dotti, J.~Oliva and R.~Troncoso,
  Phys.\ Rev.\ D {\bf 76} (2007) 064038
  [arXiv:0706.1830 [hep-th]].
\bibitem{Ishihara:2005dp}
  H.~Ishihara and K.~Matsuno,
  Prog.\ Theor.\ Phys.\  {\bf 116} (2006) 417
  [arXiv:hep-th/0510094].
\bibitem{McInnes:2010ti}
  B.~McInnes,
  Nucl.\ Phys.\ B {\bf 842} (2011) 86
  [arXiv:1008.0231 [hep-th]].
  
\bibitem{Brihaye:2007ju}
  Y.~Brihaye, T.~Delsate and E.~Radu,
  Phys.\ Lett.\  B {\bf 662} (2008) 264
  [arXiv:0710.4034 [hep-th]].
\bibitem{Delsate:2008kw}
  T.~Delsate,
  Phys.\ Lett.\ B {\bf 663} (2008) 118
  [arXiv:0802.1392 [hep-th]].
  
\bibitem{Chamseddine:1999xk}
  A.~H.~Chamseddine and W.~A.~Sabra,
  Phys.\ Lett.\ B {\bf 477}, 329 (2000)
  [arXiv:hep-th/9911195];
  \\
  D.~Klemm and W.~A.~Sabra,
  Phys.\ Rev.\ D {\bf 62}, 024003 (2000)
  [arXiv:hep-th/0001131];
  \\
  W.~A.~Sabra,
  Phys.\ Lett.\ B {\bf 545}, 175 (2002)
  [arXiv:hep-th/0207128].
\bibitem{Brihaye:2007vm}
  Y.~Brihaye, E.~Radu and C.~Stelea,
  Class.\ Quant.\ Grav.\  {\bf 24} (2007) 4839
  [arXiv:hep-th/0703046].
 \bibitem{Brihaye:2007jua}
  Y.~Brihaye and E.~Radu,
  Phys.\ Lett.\  B {\bf 658} (2008) 164
  [arXiv:0706.4378 [hep-th]].
\bibitem{Bernamonti:2007bu}
  A.~Bernamonti, M.~M.~Caldarelli, D.~Klemm, R.~Olea, C.~Sieg and E.~Zorzan,
  JHEP {\bf 0801} (2008) 061
  [arXiv:0708.2402 [hep-th]].
\bibitem{Gregory:1993vy}
  R.~Gregory and R.~Laflamme,
  Phys.\ Rev.\ Lett.\  {\bf 70} (1993) 2837
  [arXiv:hep-th/9301052].
\bibitem{Delsate:2009bd}
  T.~Delsate,
  JHEP {\bf 0907} (2009) 035
  [arXiv:0904.2149 [hep-th]].
\bibitem{Anabalon:2009kq}
  A.~Anabalon, N.~Deruelle, Y.~Morisawa, J.~Oliva, M.~Sasaki, D.~Tempo and R.~Troncoso,
  Class.\ Quant.\ Grav.\  {\bf 26} (2009) 065002
  [arXiv:0812.3194 [hep-th]].
 \bibitem{Kunz:2005nm}
  J.~Kunz, F.~Navarro-Lerida and A.~K.~Petersen,
  Phys.\ Lett.\ B {\bf 614} (2005) 104
  [arXiv:gr-qc/0503010].
\bibitem{Brihaye:2010wx}
  Y.~Brihaye, B.~Kleihaus, J.~Kunz and E.~Radu,
  JHEP {\bf 1011} (2010) 098
  [arXiv:1010.0860 [hep-th]].
\bibitem{Hawking:1998kw}
  S.~W.~Hawking, C.~J.~Hunter and M.~Taylor,
  Phys.\ Rev.\ D {\bf 59} (1999) 064005
  [hep-th/9811056].
\bibitem{COLSYS}
 U. Ascher, J. Christiansen, R.~D. Russell,
 Mathematics of Computation {\bf 33} (1979) 659;
\\
 U. Ascher, J. Christiansen, R.~D. Russell,
 ACM Transactions {\bf 7} (1981) 209.  
 

\bibitem{Kunduri:2007qy}
  H.~K.~Kunduri and J.~Lucietti,
  JHEP {\bf 0712} (2007) 015
  [arXiv:0708.3695 [hep-th]].
\bibitem{Astefanesei:2006dd}
  D.~Astefanesei, K.~Goldstein, R.~P.~Jena, A.~Sen and S.~P.~Trivedi,
  JHEP {\bf 0610} (2006) 058
  [arXiv:hep-th/0606244].
\bibitem{Brihaye:2012ww}
  Y.~Brihaye, B.~Hartmann and S.~Tojiev,
  Phys.\ Rev.\ D {\bf 87} (2013) 024040
  [arXiv:1210.2268 [gr-qc]].
 
 \end{thebibliography}
\end{document}